\documentclass[aps, pre, reprint, amsmath, amssymb]{revtex4-2}

\usepackage{graphicx}
\usepackage{bm}

\usepackage{textcomp,marvosym}
\usepackage{float}

\usepackage{bm}
\usepackage{ulem}
\usepackage{xcolor}
\usepackage{algorithm}
\usepackage{algpseudocode}
\usepackage{stmaryrd} 

\begin{document}

\title{
Phase reduction analysis of traveling breathers in reaction--diffusion systems
}

\author{Takahiro Arai}
\email{araitak@jamstec.go.jp}
\affiliation{Center for Mathematical Science and Advanced Technology, Japan Agency for Marine-Earth Science and Technology, Yokohama 236-0001, Japan}

\author{Yoji Kawamura}
\affiliation{Center for Mathematical Science and Advanced Technology, Japan Agency for Marine-Earth Science and Technology, Yokohama 236-0001, Japan}

\date{\today}

\clearpage
\begin{abstract}
We formulate a theory for phase reduction analysis of traveling breathers in reaction--diffusion systems with spatial translational symmetry.
In this formulation, the spatial and temporal phases represent the position and oscillation of a traveling breather, respectively.
We perform phase reduction analysis on a pair of FitzHugh--Nagumo models exhibiting standing breathers and a pair of Gray--Scott models exhibiting traveling breathers.  
The derived phase equations for the spatial and temporal phases indicate nontrivial spatiotemporal dynamics, where both phases are mutually coupled.
Using the phase equations, we obtain the time evolution of the phase differences, which is consistent with that obtained from direct numerical simulations.
\end{abstract}

\maketitle

\clearpage
\section{INTRODUCTION}
\label{sec:intro}

\par
Synchronization phenomena are usually important in various fields, including physics, chemistry, and life sciences~\cite{kuramoto_chemical_1984,winfree_geometry_1980,pikovsky_synchronization_2001,manrubia_emergence_2004,osipov_synchronization_2007}.
Rhythmic systems are typically described as self-sustained oscillators, where ordinary differential equations have stable limit-cycle solutions.
In such systems, phase reduction theory provides a useful framework for elucidating the synchronization properties of coupled oscillators~\cite{kuramoto_chemical_1984,winfree_geometry_1980,pikovsky_synchronization_2001,manrubia_emergence_2004,osipov_synchronization_2007,brown_phase_2004,ermentrout_type_1996,nakao_phase_2016,pietras_network_2019,ermentrout_mathematical_2010,hoppensteadt_weakly_1997,izhikevich_dynamical_2006}. 
On the basis of this framework, the collective synchronization of phase oscillators has been extensively investigated in various network configurations including globally coupled systems, nonlocally coupled systems, and complex networks ~\cite{kuramoto_chemical_1984,winfree_geometry_1980,pikovsky_synchronization_2001,manrubia_emergence_2004,osipov_synchronization_2007,kuramoto_chemical_1984,winfree_geometry_1980,pikovsky_synchronization_2001,manrubia_emergence_2004,osipov_synchronization_2007,brown_phase_2004,ermentrout_type_1996,nakao_phase_2016,pietras_network_2019,ermentrout_mathematical_2010,hoppensteadt_weakly_1997,izhikevich_dynamical_2006,acebron_kuramoto_2005,arenas_synchronization_2008,ashwin_mathematical_2016,pikovsky_dynamics_2015,stankovski_coupling_2017,strogatz_kuramoto_2000,rodrigues_kuramoto_2016}.
Furthermore, data-driven approaches based on the phase reduction theory have enabled us to uncover synchronization properties in various phenomena.  
For example, such approaches have been employed in the fields of neurophysiology and biology, for the locomotor coordination of humans and millipedes~\cite{arai_interlimb_2024,furukawa_bayesian_2025,funato_evaluation_2016,yoshikawa_wavy_nodate}, brain waves~\cite{stankovski_coupling_2017,onojima_dynamical_2018}, the human cardiorespiratory system~\cite{kralemann_vivo_2013}, and selective attention mechanisms in frog choruses~\cite{ota_interaction_2020}.

\par
The theory for phase reduction analysis has been generalized to spatiotemporal dynamics.
Recent studies have developed phase reduction theory for limit-cycle solutions to partial differential equations describing oscillatory thermal convection~\cite{kawamura_collective_2013,kawamura_noise-induced_2014}, periodic flows~\cite{godavarthi_optimal_2023,kawamura_adjoint-based_2022,taira_phase-response_2018,iima_optimal_2024,iima_phase_2021,iima_jacobian-free_2019}, and beating flagella~\cite{kawamura_phase_2018}. 
Furthermore, considering that the traveling solution~\cite{kawamura_collective_2008, kawamura_phase_2010} and oscillating solution~\cite{kawamura_collective_2011, kawamura_collective_2017} of the nonlinear Fokker--Planck equations are within the scope of the phase reduction theory, collective oscillations of globally coupled elements can be analyzed.
In reaction--diffusion systems, which are particularly important in this study, theoretical frameworks have also been developed for traveling pulses~\cite{lober_controlling_2014,lober_front_2012} and rotating spirals~\cite{biktashev_orbital_2010,sandstede_dynamics_1997}.
Notably, in Refs.~\cite{nakao_phase-reduction_2014,kawamura_optimizing_2017}, phase equations can be derived without assuming symmetry with respect to continuous spatial translation or rotation, making the theory applicable to each of traveling solution, rotating solution, and oscillating solution.
Developing a phase reduction theory for spatiotemporal patterns is essential for understanding and controlling various experimental systems, such as the Belousov--Zhabotinsky reaction~\cite{hildebrand_synchronization_2003,khan_complete_2024,kalita_rotational_2022,khan_synchronization_2024}.

\par 
For reaction--diffusion systems, the synchronization properties of spatiotemporal dynamics can be investigated using phase reduction theory, which can be applied to partial differential equations with limit-cycle solutions; however, its applicability is limited to specific types of spatiotemporal patterns.
For example, the theory developed in Ref.~\cite{nakao_phase-reduction_2014} can be applied to circulating pulses, rotating spirals, oscillating spots, and target waves.  
These patterns typically rely on a single symmetry, such as spatial translational, spatial rotational, or temporal translational symmetry.
In contrast, patterns involving both spatial and temporal translational symmetries are outside the scope of the theory.

\par 
Traveling breather and standing breather~\cite{yadome_chaotic_2011,hagberg_pattern_1994}, which are typical examples of traveling and oscillating spatiotemporal patterns, cannot be analyzed using the existing theory for the limit-cycle solution.
A traveling breather spontaneously breaks spatial reflection symmetry of the system; as a result, its traveling velocity is nonzero. 
A standing breather can be regarded as a special case, where spatial reflection symmetry is not broken.
For these breathers, the formulation of a theory for the phase reduction analysis describing the mutual influences between translation and oscillation is necessary.

\par 
Recent developments in phase reduction theory for oscillatory thermal convection systems with spatial translational symmetry~\cite{kawamura_phase_description_2015,kawamura_phase_2019} provide a foundation for analyzing the synchronization properties of breathers. 
In these studies, traveling and oscillating spatiotemporal patterns of oscillatory thermal convection were investigated in a cylindrical Hele--Shaw cell and a two-dimensional incompressible Navier--Stokes flow system.
Both systems exhibit lateral periodicity and are governed by partial differential equations with limit-torus solutions.
The theory proposed in Refs.~\cite{kawamura_phase_description_2015,kawamura_phase_2019} are a generalization of the theory from partial differential equations with limit-cycle solutions~\cite{nakao_phase-reduction_2014, kawamura_collective_2011, kawamura_collective_2013} to those with limit-torus solutions.

\par
In this study, we formulate a phase reduction theory for traveling and oscillating solutions in reaction--diffusion systems with spatial translational symmetry. 
The formulation of the theory is based on Refs.~\cite{kawamura_phase_description_2015,kawamura_phase_2019}.
Then, we apply the theory to a pair of FitzHugh--Nagumo (FHN) models exhibiting standing breathers~\cite{hagberg_pattern_1994} and a pair of Gray--Scott models exhibiting traveling breathers~\cite{yadome_chaotic_2011}, and investigate their synchronization properties.
The theory formulated in this study is applicable to both traveling breathers and standing breathers; the latter correspond to the case of zero traveling velocity.

\par
\begin{figure*}[h]
    \begin{center}
        \includegraphics[scale=0.9]{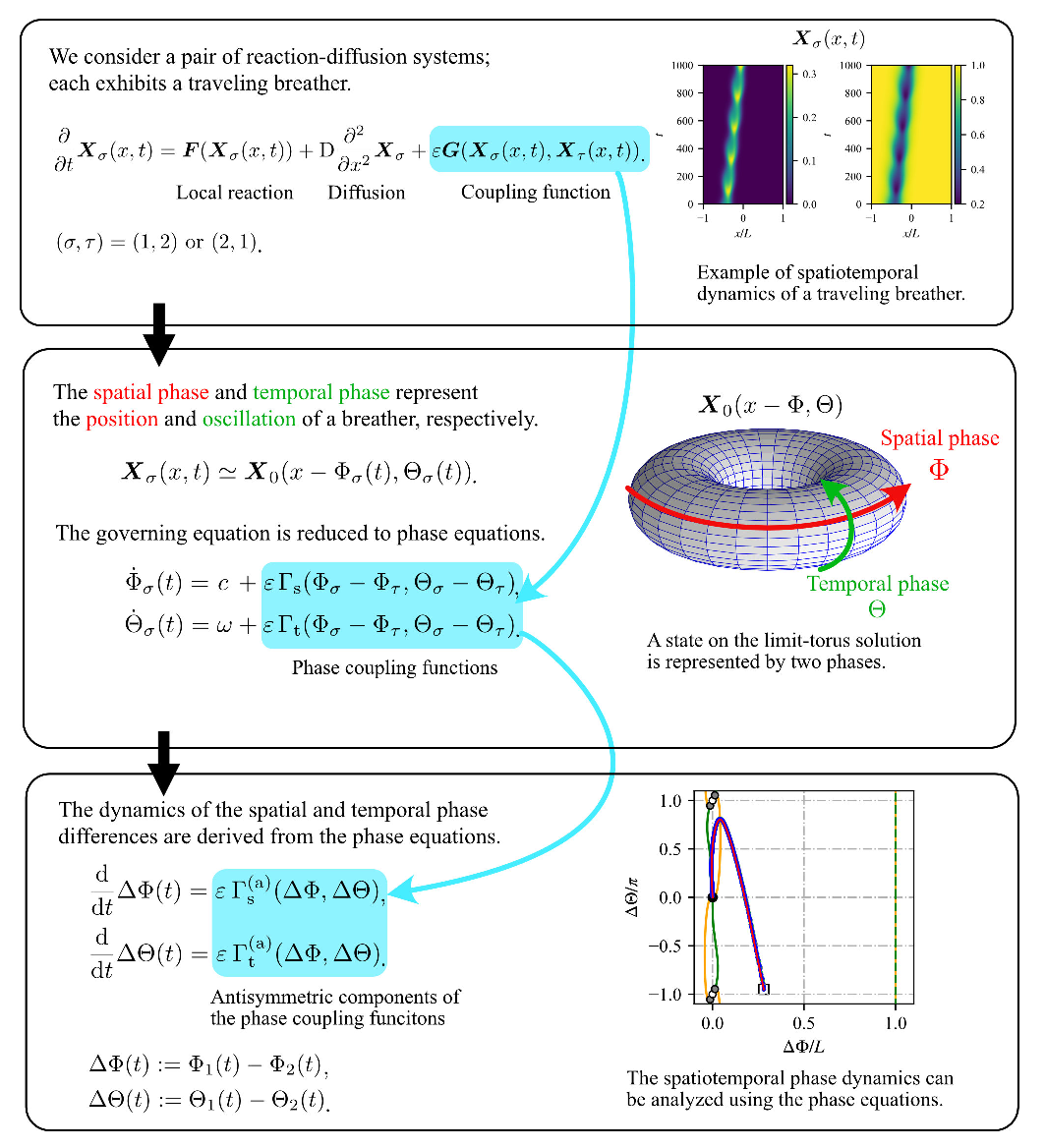}
        \caption{
        Overview of this study. 
        (top)~
        A pair of reaction--diffusion systems exhibiting traveling breathers. 
        The systems satisfy the homogeneity of the medium and periodic boundary conditions.
        The coupling function between the systems affects the dynamics of the breather, leading to variations in its traveling velocity and oscillation frequency.
        (middle)~
        Phase equations representing the spatiotemporal phase dynamics of the traveling breathers.
        The limit-torus solution is represented by a spatial phase and a temporal phase, corresponding to the position and oscillation of the breather, respectively.
        The phase equations for both phases are derived from the governing equations of the reaction--diffusion systems using phase reduction analysis.
        The phase coupling functions in the phase equations describe the effect of coupling between the systems on the phases.
        (bottom)~
        Analysis of synchronization properties.  
        The dynamics of the spatial and temporal phase differences are derived from the phase equations shown in the middle panel.
        By analyzing these dynamics, the time evolution of the phase differences and the stability of the synchronized states between breathers can be investigated.
        The figure in the bottom panel shows the result, including a comparison between theoretical values and direct numerical simulations.
        Additionally, it shows the nullclines and fixed points corresponding to the synchronized states as well as their linear stability.
        Details of this figure are provided in Sec.~\ref{subsec:coupled_Gray--Scott_Model} and Fig.~\ref{fig:fig14}(e).
        }
        \label{fig:fig1}
    \end{center}
\end{figure*}

 
An overview of the present study is shown in Fig.~\ref{fig:fig1}.
We consider reaction--diffusion systems exhibiting traveling and oscillating spatiotemporal patterns, i.e., traveling breathers (top panel in Fig.~\ref{fig:fig1}).  
The system satisfies medium homogeneity and periodic boundary conditions and therefore exhibits spatial translational symmetry.
The coupling function of a coupled system affects its breather dynamics, leading to variations in the traveling velocity and oscillation frequency of the breather.
On the basis of the phase reduction theory formulated in this study, we derive the spatiotemporal phase dynamics of the system as phase equations (middle panel in Fig.~\ref{fig:fig1}).
The position and oscillation of a breather are represented by the spatial phase and temporal phase, respectively.
The temporal phase is a concept similar to that used in conventional theory for limit-cycle solutions, while the spatial phase is an additional concept arising from the spatial translational symmetry.
The phase coupling functions in the phase equations describe the effect of coupling between a pair of reaction--diffusion systems on the spatial and temporal phases.
Using the phase equations, we obtain the theoretical values of spatiotemporal phase dynamics (bottom panel in Fig.~\ref{fig:fig1}).  
On the basis of the phase equations, we derive the dynamics of spatial and temporal phase differences. 
The analysis of these dynamics enables us to investigate the time evolution of spatial and temporal phase differences as well as the stability of the synchronized states between breathers. 
Furthermore, we validate the theoretical values by comparing the time evolution of the phase differences obtained from the phase equations with those obtained from direct numerical simulations.

\par
This paper is organized as follows.
In Sec.~\ref{sec:Theory}, we formulate a theory for the phase reduction analysis of a reaction--diffusion system exhibiting a traveling breather.
In Sec.~\ref{sec:NumericalAnalysis}, we analyze the spatiotemporal phase dynamics of a pair of FHN models exhibiting standing breathers (Sec.~\ref{subsec:coupled_FHN}) and a pair of Gray--Scott models exhibiting traveling breathers (Sec.~\ref{subsec:coupled_Gray--Scott_Model}).
Concluding remarks are given in Sec.~\ref{sec:concluding_remarks}.
Supplementary information on phase reduction analysis is provided in Appendices~\ref{appendix:A}--\ref{appendix:C}.

\section{THEORY FOR PHASE REDUCTION ANALYSIS OF TRAVELING BREATHERS}
\label{sec:Theory}

\par
In this section, we formulate a theory for phase reduction analysis of reaction--diffusion systems exhibiting traveling breathers. 
The proposed theory is based on the framework developed in Refs.~\cite{kawamura_phase_description_2015,kawamura_phase_2019}.

\subsection{Reaction--diffusion system with spatial translational symmetry}
\label{subsec:RD-system}

\par
We consider one-dimensional reaction--diffusion systems, which are described by the following partial differential equation:
\begin{align}
    \label{eq:RD-systems}
    \frac{\partial}{\partial t} \boldsymbol{X}(x,t) = \boldsymbol{F}(\boldsymbol{X}(x,t)) + {\rm D} \frac{\partial^2}{\partial x^2} \boldsymbol{X}, 
\end{align}
where the vector field $\boldsymbol{X}(x,t) \in \mathbb{R}^N$ denotes a state of the medium at point $x$ at time $t$, $\boldsymbol{F}(\boldsymbol{X}(x,t)) \in \mathbb{R}^N$ denotes the local reaction dynamics, and ${\rm D} (\partial^2 \boldsymbol{X} /\partial x^2)$ denotes the diffusion of $\boldsymbol{X}(x,t)$ over the medium with a diffusion coefficient matrix ${\rm D} \in \mathbb{R}^{N \times N}$.
The system size is $2L$, i.e., $x \in [0, 2L)$. 
We consider the $2L$-periodic boundary conditions as follows:
\begin{align}
\label{eq:periodic_condition}
\boldsymbol{X}(x + 2L, t) = \boldsymbol{X}(x, t).
\end{align}
Because of the homogeneity of the medium in Eq.~(\ref{eq:RD-systems}) and the periodic boundary condition on $x$, the system exhibits continuous spatial translational symmetry with respect to $x$.

\par
Furthermore, the system exhibits spatial reflection symmetry, meaning that Eq.~(\ref{eq:RD-systems}) is invariant under the transformation $x \to -x$. 
However, the proposed theory does not require this reflection symmetry.

\subsection{Limit-torus solutions and their Floquet-type systems}

\par 
We assume that the reaction--diffusion system (Eq.~(\ref{eq:RD-systems})) has a stable limit-torus solution, representing a traveling and oscillating spatiotemporal pattern, i.e., a traveling breather.
The limit-torus solution is described as follows:
\begin{align}
\label{eq:limit-torus}
& \boldsymbol{X}(x,t) = \boldsymbol{X}_0(x - \Phi(t), \Theta(t)), ~
\dot{\Phi}(t) = c, ~
\dot{\Theta}(t) = \omega.
\end{align}
The position and oscillation of the spatiotemporal pattern are represented by the spatial phase $\Phi \in [0, 2L)$ and temporal phase $\Theta \in [0, 2\pi)$, respectively. 
The traveling velocity $c$ and oscillation frequency $\omega$ are constant.
A breather with $c \neq 0$ is referred to as a traveling breather, whereas a breather with $c = 0$ is referred to as a standing breather~\cite{yadome_chaotic_2011}. 
The standing breather is regarded as a special case of the traveling breather.
The limit-torus solution $\boldsymbol{X}_0(x - \Phi, \Theta)$ is $2L$-periodic and $2\pi$-periodic with respect to $(x-\Phi)$ and $\Theta$, respectively, i.e., 
\begin{align}
\label{eq:periodic_condition_X0}
& \boldsymbol{X}_0(x-\Phi+2L, \Theta) = \boldsymbol{X}_0(x-\Phi, \Theta), \\
& \boldsymbol{X}_0(x-\Phi, \Theta+2\pi) = \boldsymbol{X}_0(x-\Phi, \Theta).
\end{align}
By substituting Eq.~(\ref{eq:limit-torus}) into Eq.~(\ref{eq:RD-systems}), we obtain that $\boldsymbol{X}_0(x-\Phi, \Theta)$ satisfies the following equation:
\begin{align}
\label{eq:limit-torus-eq}
& \left[ -c  \frac{\partial}{\partial x} + \omega \frac{\partial}{\partial \Theta} \right] \boldsymbol{X}_0 \left( x - \Phi, \Theta \right) 
\nonumber \\
& = \boldsymbol{F}\left(\boldsymbol{X}_0 (x-\Phi, \Theta) \right) + D \frac{\partial^2}{\partial x^2} \boldsymbol{X}_0.
\end{align}

\par 
Let us consider a small deviation $\boldsymbol{y}(x - \Phi, \Theta, t) \in \mathbb{R}^N$ from the limit-torus solution $\boldsymbol{X}_0(x-\Phi, \Theta)$. 
A perturbed solution can be expressed as 
\begin{align}
\label{eq:deviation_y}
& \boldsymbol{X}(x, t) = \boldsymbol{X}_0 \left( x - \Phi, \Theta \right) + \boldsymbol{y} \left( x - \Phi, \Theta, t \right).
\end{align}
By linearizing of Eq.~(\ref{eq:RD-systems}) with respect to $\boldsymbol{y}(x - \Phi, \Theta, t)$, we obtain the following equation:
\begin{align}
\label{eq:time_evolution_y}
& \frac{\partial}{\partial t} \boldsymbol{y}(x - \Phi, \Theta, t) 
= \hat{\mathcal{L}}(x - \Phi, \Theta)  \boldsymbol{y}(x - \Phi, \Theta, t).
\end{align}
The linear operator in Eq.~(\ref{eq:time_evolution_y}) is defined as follows:
\begin{align}
\label{eq:linear_operator}
&  \hat{\mathcal{L}}(x - \Phi, \Theta) \boldsymbol{y} \left( x - \Phi, \Theta\right) \nonumber \\
& = \left[ \hat{L}(x-\Phi, \Theta) + c \frac{\partial}{\partial x}
- \omega \frac{\partial}{\partial \Theta} \right]
\boldsymbol{y} \left( x - \Phi, \Theta \right), 
\end{align}
where 
\begin{align}
\label{eq:linear_operator_sub}
& \hat{L} (x-\Phi, \Theta) \bm{y}(x-\Phi, \Theta)  \nonumber \\
& := \left[ {\rm J}(\boldsymbol{X}_0( x - \Phi, \Theta)) + {\rm D} \frac{\partial^2}{\partial x^2} \right] \boldsymbol{y} \left( x - \Phi, \Theta \right).
\end{align}
The Jacobi matrix of $\boldsymbol{F}(\boldsymbol{X}(x,t))$ at $\boldsymbol{X}(x,t) = \boldsymbol{X}_0(x-\Phi, \Theta)$ is denoted by ${\rm J}(\boldsymbol{X}_0( x - \Phi, \Theta)) \in \mathbb{R}^{N \times N}$.
In Eqs.~(\ref{eq:linear_operator}) and (\ref{eq:linear_operator_sub}), we omit the explicit $t$-dependence of $\boldsymbol{y}(x - \Phi, \Theta, t)$ and simply denote it as $\boldsymbol{y}(x - \Phi, \Theta)$.
Note that the $t$-dependence of function $\boldsymbol{y}(x-\Phi, \Theta)$ does not appear hereafter since we consider only the eigenvalue problem of the linear operator $\hat{\mathcal{L}}(x - \Phi, \Theta)$.
As in the limit-torus solution $\boldsymbol{X}_0(x-\Phi, \Theta)$, function $\boldsymbol{y}(x-\Phi, \Theta)$ satisfies the $2L$-periodicity and $2\pi$-periodicity with respect to $(x-\Phi)$ and $\Theta$, respectively.
Note that the linear operator $\hat{\mathcal{L}}(x - \Phi, \Theta)$ is periodic with respect to both $(x-\Phi)$ and $\Theta$.
Therefore, Eq.~(\ref{eq:time_evolution_y}) represents a Floquet-type system with two zero eigenvalues.  
In such systems, the two zero eigenvalues are associated with breaking of spatial and temporal translational symmetries.

\par
We introduce the adjoint operator of the linear operator $\hat{\mathcal{L}}(x - \Phi, \Theta)$ by defining the inner product of two functions as follows:
\begin{align}
\label{eq:inner_product}
& \left\llbracket \boldsymbol{y}^*(x - \Phi, \Theta),  \boldsymbol{y}(x - \Phi, \Theta) \right\rrbracket 
\nonumber \\ 
& := \frac{1}{2\pi} \int_0^{2\pi} \mathrm{d}\Theta\, \int_0^{2L} \mathrm{d}x\, 
\boldsymbol{y}^*(x - \Phi, \Theta) \cdot \boldsymbol{y}(x - \Phi, \Theta).
\end{align}
The definition of the adjoint operator $\hat{\mathcal{L}}^*(x - \Phi, \Theta)$ is based on the following relation:
\begin{align}
\label{eq:adjoint_relationship}
& \left\llbracket \boldsymbol{y}^*(x - \Phi, \Theta), 
\hat{\mathcal{L}}(x - \Phi, \Theta)  \boldsymbol{y}(x - \Phi, \Theta) \right\rrbracket \nonumber \\
& = \left\llbracket \hat{\mathcal{L}}^*(x - \Phi, \Theta)  \boldsymbol{y}^*(x - \Phi, \Theta), 
\boldsymbol{y}(x - \Phi, \Theta) \right\rrbracket.
\end{align}
By partially integrating Eq.~(\ref{eq:adjoint_relationship}), the adjoint operator $\hat{\mathcal{L}}^*(x - \Phi, \Theta)$ is explicitly described as
\begin{align}
\label{eq:linear_operator*}
& \hat{\mathcal{L}}^*(x - \Phi, \Theta)  \boldsymbol{y}^*(x-\Phi, \Theta) \nonumber \\ 
& = \left[ \hat{{\rm L}}^*(x-\Phi, \Theta) - c \frac{\partial}{\partial x} + \omega \frac{\partial}{\partial \Theta} \right] \boldsymbol{y}^*(x-\Phi, \Theta),
\end{align}
where
\begin{align}
\label{eq:lineqr_operator_sub*}
& \hat{{\rm L}}^*(x-\Phi, \Theta) \boldsymbol{y}^*(x-\Phi, \Theta) \nonumber \\
& := \left[ {\rm J}^{\mathsf{T}}(\boldsymbol{X}_0(x-\Phi, \Theta)) + {\rm D}^{\mathsf{T}} \frac{\partial^2}{\partial x^2} \right] \boldsymbol{y}^*(x-\Phi, \Theta).
\end{align}
In the derivation of Eqs.~(\ref{eq:linear_operator*}) and (\ref{eq:lineqr_operator_sub*}), we assumed that $\boldsymbol{y}^*(x - \Phi, \Theta)$ satisfies the periodic boundary conditions with a period $2L$ in $(x - \Phi)$ and $2\pi$ in $\Theta$; these serve as adjoint boundary conditions.  
As a result, the bilinear concomitant in Eq.~(\ref{eq:adjoint_relationship}) vanishes.

\subsection{Floquet zero eigenfunctions and their orthonormalization}

\par 
We use the Floquet and adjoint eigenfunctions associated
with the two zero eigenvalues of $\hat{\mathcal{L}}(x - \Phi, \Theta)$ and $\hat{\mathcal{L}}^*(x - \Phi, \Theta)$.
According to Eqs.~(\ref{eq:linear_operator}) and (\ref{eq:linear_operator*}), these eigenfunctions satisfy the following equations:
\begin{align}
\label{eq:Floquet_zero_eigengunction_Us}
& \hat{\mathcal{L}}(x - \Phi, \Theta)  \boldsymbol{U}_{\mathrm{s}}(x - \Phi, \Theta) \nonumber \\
&= \left[ \hat{{\rm L}}(x-\Phi, \Theta) + c \frac{\partial}{\partial x} - \omega \frac{\partial}{\partial \Theta} \right]
\boldsymbol{U}_{\mathrm{s}}(x - \Phi, \Theta) = 0,  \\
\label{eq:Floquet_zero_eigengunction_Ut}
& \hat{\mathcal{L}}(x - \Phi, \Theta)  \boldsymbol{U}_{\mathrm{t}}(x - \Phi, \Theta) \nonumber \\
&= \left[ \hat{{\rm L}}(x-\Phi, \Theta) + c \frac{\partial}{\partial x} - \omega \frac{\partial}{\partial \Theta} \right]
\boldsymbol{U}_{\mathrm{t}}(x - \Phi, \Theta) = 0,  \\
\label{eq:adjoint_zero_eigengunction_Us*}
& \hat{\mathcal{L}}^*(x - \Phi, \Theta)  \boldsymbol{U}_{\mathrm{s}}^*(x - \Phi, \Theta) \nonumber \\
&= \left[ \hat{{\rm L}}^*(x-\Phi, \Theta) - c \frac{\partial}{\partial x} + \omega \frac{\partial}{\partial \Theta} \right]
\boldsymbol{U}_{\mathrm{s}}^*(x - \Phi, \Theta) = 0, \\
\label{eq:adjoint_zero_eigengunction_Ut*}
& \hat{\mathcal{L}}^*(x - \Phi, \Theta)  \boldsymbol{U}_{\mathrm{t}}^*(x - \Phi, \Theta) \nonumber \\
&= \left[ \hat{{\rm L}}^*(x-\Phi, \Theta) - c \frac{\partial}{\partial x} + \omega \frac{\partial}{\partial \Theta} \right]
\boldsymbol{U}_{\mathrm{t}}^*(x - \Phi, \Theta) = 0, 
\end{align}
where $\boldsymbol{U}_\mathrm{p}(x - \Phi, \Theta) \in \mathbb{R}^N$ and $\boldsymbol{U}_\mathrm{p}^*(x - \Phi, \Theta) \in \mathbb{R}^N$ ($\mathrm{p} \in \{\mathrm{s}, \mathrm{t}\}$) denote the Floquet zero eigenfunction and its corresponding adjoint zero eigenfunctions, respectively.
The spatial phase is associated with $\boldsymbol{U}_\mathrm{s}(x-\Phi, \Theta)$ and $\boldsymbol{U}_\mathrm{s}^*(x-\Phi, \Theta)$, whereas the temporal phase is associated with $\boldsymbol{U}_\mathrm{t}(x-\Phi, \Theta)$ and $\boldsymbol{U}_\mathrm{t}^*(x-\Phi, \Theta)$.
The two Floquet zero eigenfunctions can be selected as 
\begin{align}
\label{eq:choice_of_Us}
\boldsymbol{U}_\mathrm{s}(x - \Phi, \Theta)
&= \frac{\partial}{\partial x} \boldsymbol{X}_0(x - \Phi, \Theta), \\
\label{eq:choice_of_Ut}
\boldsymbol{U}_\mathrm{t}(x - \Phi, \Theta)
&= \frac{\partial}{\partial \Theta} \boldsymbol{X}_0(x - \Phi, \Theta).
\end{align}
The selected eigenfunctions can be verified by differentiating Eq.~(\ref{eq:limit-torus-eq}) with respect to $x$ and $\Theta$, respectively.
For the inner product defined in Eq.~(\ref{eq:inner_product}) and the two Floquet zero eigenfunctions given in Eqs.~(\ref{eq:choice_of_Us}) and (\ref{eq:choice_of_Ut}), the corresponding adjoint zero eigenfunctions satisfy the following orthonormalization condition:
\begin{align}
\label{eq:orthonormalization}
& \llbracket  \boldsymbol{U}_\mathrm{p}^*(x - \Phi, \Theta), 
\boldsymbol{U}_\mathrm{q}(x - \Phi, \Theta)  \rrbracket   
\nonumber \\
&= \frac{1}{2\pi} \int_0^{2\pi} \mathrm{d}\Theta\, \int_0^{2L} \mathrm{d}x\,
\boldsymbol{U}_\mathrm{p}^*(x - \Phi, \Theta) \cdot \boldsymbol{U}_\mathrm{q}(x - \Phi, \Theta) 
= \delta_{\mathrm{pq}},
\end{align}
where $\mathrm{p}, \mathrm{q} \in \{\mathrm{s}, \mathrm{t}\}$ and $\delta_{\mathrm{pq}}$ denotes the Kronecker delta.
Additionally, the following condition is satisfied:
\begin{align}
\label{eq:theta_diff_orthonarmalization}
& \frac{\partial}{\partial \Theta} \int_0^{2L} \mathrm{d}x\,  
        \boldsymbol{U}_\mathrm{p}^*(x - \Phi, \Theta) \cdot 
       \boldsymbol{U}_\mathrm{q}(x - \Phi, \Theta) 
\nonumber \\
&= \int_0^{2L} \mathrm{d}x\,
\Biggl\{
    \boldsymbol{U}_\mathrm{p}^*(x - \Phi, \Theta) \cdot 
    \left( \frac{\partial}{\partial \Theta} 
            \boldsymbol{U}_\mathrm{q}(x - \Phi, \Theta) \right)
    \nonumber \\
    &\qquad\quad +
    \left( \frac{\partial}{\partial \Theta} 
            \boldsymbol{U}_\mathrm{p}^*(x - \Phi, \Theta) \right) 
    \cdot \boldsymbol{U}_\mathrm{q}(x - \Phi, \Theta) 
\Biggr\}
\nonumber \\
&= \frac{1}{\omega} \int_0^{2L} \mathrm{d}x\,
\Biggl\{
    \boldsymbol{U}_\mathrm{p}^*(x - \Phi, \Theta) 
    \nonumber \\
    & \qquad\quad \cdot 
    \left[ 
        \hat{{\rm L}}(x - \Phi, \Theta) + c \frac{\partial}{\partial x} 
    \right] 
    \boldsymbol{U}_\mathrm{q}(x - \Phi, \Theta) 
    \nonumber \\
    & \qquad\quad
    - \left[ 
        \hat{{\rm L}}^*(x - \Phi, \Theta) - c \frac{\partial}{\partial x} 
    \right] 
    \boldsymbol{U}_\mathrm{p}^*(x - \Phi, \Theta)  
    \nonumber \\
    & \qquad\quad 
    \cdot \boldsymbol{U}_\mathrm{q}(x - \Phi, \Theta) 
\Biggr\} = 0,
\end{align}
where Eqs.~(\ref{eq:Floquet_zero_eigengunction_Us})--(\ref{eq:adjoint_zero_eigengunction_Ut*}) are employed in the derivation.
From Eq.~(\ref{eq:theta_diff_orthonarmalization}), the following orthonormalization condition is satisfied independently for each $\Theta$:
\begin{align}
\label{eq:orthonormalization_independent_of_theta}
& \int_0^{2L} \mathrm{d}x\, \boldsymbol{U}_\mathrm{p}^*(x - \Phi, \Theta) \cdot \boldsymbol{U}_\mathrm{q}(x - \Phi, \Theta) = \delta_{\mathrm{pq}}.
\end{align}

\par 
According to Eqs.~(\ref{eq:adjoint_zero_eigengunction_Us*}) and (\ref{eq:adjoint_zero_eigengunction_Ut*}), we obtain the following adjoint equation:
\begin{align}
\label{eq:adjoint_equation}
& \omega \frac{\partial}{\partial \Theta} \boldsymbol{U}_\mathrm{p}^*(x - \Phi, \Theta) \nonumber \\
& = - \left[ \hat{{\rm L}}^* (x-\Phi, \Theta) - c \frac{\partial}{\partial x}  \right] 
\boldsymbol{U}_\mathrm{p}^*(x - \Phi, \Theta), 
\end{align}
for $\mathrm{p} \in \{\mathrm{s}, \mathrm{t}\}$.
By substituting $\Theta = - \omega s$, the above equation is transformed into the following equation:
\begin{align}
\label{eq:adjoint_equation_2}
& \frac{\partial}{\partial s} 
\boldsymbol{U}_\mathrm{p}^*(x - \Phi, -\omega s) \nonumber \\
& = \left[ \hat{{\rm L}}^* (x-\Phi, -\omega s) - c \frac{\partial}{\partial x}
\right]
\boldsymbol{U}_\mathrm{p}^*(x - \Phi, -\omega s).
\end{align}
A relaxation method based on the adjoint equation~(\ref{eq:adjoint_equation_2}), which is also referred to as the adjoint method, is employed to obtain the adjoint zero eigenfunctions.
The details of the adjoint method and the orthonormalization method, which is based on Eq.~(\ref{eq:orthonormalization}), are described in Appendix~\ref{appendix:A}.
The adjoint method has also been used in previous studies for limit-cycle solutions to ordinary differential equations~\cite{brown_phase_2004,ermentrout_type_1996,nakao_phase_2016,pietras_network_2019, ermentrout_mathematical_2010,hoppensteadt_weakly_1997,izhikevich_dynamical_2006} and partial differential equations~\cite{kawamura_collective_2017,kawamura_phase_2018,nakao_phase-reduction_2014,kawamura_noise-induced_2014,kawamura_collective_2013,kawamura_collective_2011,kawamura_optimizing_2017}.
The adjoint methods presented in this study and in Refs.~\cite{kawamura_phase_2019,kawamura_phase_description_2015} can be considered as a generalization of the existing methods, extending the framework from ordinary differential equations to partial differential equations, and from limit-cycle solutions to limit-torus solutions.

\subsection{Traveling breather with weak perturbations}

\par
We consider a reaction--diffusion system with a weak perturbation, which is described as follows:
\begin{align}
\label{eq:perturbed_system}
\frac{\partial}{\partial t} \boldsymbol{X}(x, t)
= \boldsymbol{F}(\boldsymbol{X}(x,t)) + {\rm D} \frac{\partial^2}{\partial x^2} \boldsymbol{X}
+ \varepsilon  \boldsymbol{p}(x, t),
\end{align}
where $\varepsilon \boldsymbol{p}(x,t) \in \mathbb{R}^N$, with the intensity $ \varepsilon \ll 1 $, represents a weak perturbation. 
We assume that the perturbed solution $\boldsymbol{X}(x,t)$ remains in the vicinity of the orbit of the limit-torus solution $\boldsymbol{X}_0(x-\Phi, \Theta)$.
Considering that 
\begin{align}
& \frac{\partial}{\partial t} \boldsymbol{X}(x,t)  
\nonumber \\
& \simeq -\dot{\Phi}(t) \frac{\partial}{\partial x} \boldsymbol{X}_0(x-\Phi, \Theta) + \dot{\Theta}(t) \frac{\partial}{\partial \Theta} \boldsymbol{X}_0(x-\Phi, \Theta),  
\end{align}
the dynamics of the perturbed system (Eq.~(\ref{eq:perturbed_system})) can be projected onto the unperturbed limit-torus solution with respect to the spatial and temporal phases as follows:
\begin{align}
\label{eq:phase_equation_pertubed_Phi}
- \dot{\Phi}(t) 
\simeq & \int_0^{2L} \mathrm{d}x\,
\boldsymbol{U}_\mathrm{s}^*(x - \Phi, \Theta) 
\nonumber \\
& \cdot \left[ 
    \boldsymbol{F}(\boldsymbol{X}_0(x-\Phi, \Theta)) + {\rm D} \frac{\partial^2}{\partial x^2} \boldsymbol{X}_0 + \varepsilon \boldsymbol{p}(x, t) 
\right] 
\nonumber \\ 
= & \int_0^{2L} \mathrm{d}x\,
\boldsymbol{U}_\mathrm{s}^*(x - \Phi, \Theta)  
\nonumber \\
& \cdot \left[ 
    - c \frac{\partial}{\partial x} \boldsymbol{X}_0(x-\Phi, \Theta) + \omega \frac{\partial \boldsymbol{X}_0}{\partial \Theta} + \varepsilon \boldsymbol{p}(x, t) 
\right] 
\nonumber \\ 
= & \int_0^{2L} \mathrm{d}x\,
\boldsymbol{U}_\mathrm{s}^*(x - \Phi, \Theta) 
\nonumber \\ 
& \cdot \Bigl[ -c \boldsymbol{U}_\mathrm{s}(x - \Phi, \Theta) + \omega \boldsymbol{U}_\mathrm{t}(x - \Phi, \Theta) + \varepsilon \boldsymbol{p}(x, t) \Bigr] 
\nonumber \\
= & -c + \varepsilon \int_0^{2L} \mathrm{d}x\,
\boldsymbol{U}_\mathrm{s}^*(x - \Phi, \Theta) \cdot \boldsymbol{p}(x, t),
\end{align}
and
\begin{align}
\label{eq:phase_equation_pertubed_Theta}
\dot{\Theta}(t)
\simeq & \int_0^{2L} \mathrm{d}x\,
\boldsymbol{U}_\mathrm{t}^*(x - \Phi, \Theta)
\nonumber \\ 
& \cdot \left[ 
    \boldsymbol{F}(\boldsymbol{X}_0(x-\Phi, \Theta)) + {\rm D} \frac{\partial^2}{\partial x^2} \boldsymbol{X}_0 + \varepsilon \boldsymbol{p}(x, t) 
\right] 
\nonumber \\ 
= & \int_0^{2L} \mathrm{d}x\,
\boldsymbol{U}_\mathrm{t}^*(x - \Phi, \Theta) 
\nonumber \\ 
& \cdot \left[ 
    -c \frac{\partial}{\partial x} \boldsymbol{X}_0(x-\Phi, \Theta)
    + \omega \frac{\partial \boldsymbol{X}_0}{\partial \Theta}
    + \varepsilon \boldsymbol{p}(x, t) 
\right] 
\nonumber \\
= & \int_0^{2L} \mathrm{d}x\,
\boldsymbol{U}_\mathrm{t}^*(x - \Phi, \Theta) 
\nonumber \\
& \cdot 
\Bigl[ 
    -c \boldsymbol{U}_\mathrm{s}(x - \Phi, \Theta)
    + \omega \boldsymbol{U}_\mathrm{t}(x - \Phi, \Theta)
    + \varepsilon \boldsymbol{p}(x, t) 
\Bigr] 
\nonumber \\
= & \omega + \varepsilon \int_0^{2L} \mathrm{d}x\,
\boldsymbol{U}_\mathrm{t}^*(x - \Phi, \Theta) \cdot \boldsymbol{p}(x, t).
\end{align}
In the derivation of Eqs.~(\ref{eq:phase_equation_pertubed_Phi}) and (\ref{eq:phase_equation_pertubed_Theta}), we approximated $\boldsymbol{X}(x,t)$ by the unperturbed limit-torus solution $\boldsymbol{X}_0(x-\Phi, \Theta)$ and used Eqs.~(\ref{eq:limit-torus-eq}), (\ref{eq:choice_of_Us}), (\ref{eq:choice_of_Ut}), and (\ref{eq:orthonormalization_independent_of_theta}).
By rewriting Eqs.~(\ref{eq:phase_equation_pertubed_Phi}) and (\ref{eq:phase_equation_pertubed_Theta}) in the following form, we obtain the phase equations:
\begin{align}
\label{eq:phase_equation_pertubed_Phi_rewrite}
\dot{\Phi}(t)
&= c + \varepsilon \int_0^{2L} \mathrm{d}x\, 
\boldsymbol{Z}_\mathrm{s}(x - \Phi, \Theta) \cdot \boldsymbol{p}(x, t), \\
\dot{\Theta}(t)
\label{eq:phase_equation_pertubed_Theta_rewrite}
&= \omega + \varepsilon \int_0^{2L} \mathrm{d}x\,
\boldsymbol{Z}_\mathrm{t}(x - \Phi, \Theta) \cdot \boldsymbol{p}(x, t).
\end{align}
The phase sensitivity functions of the spatial and temporal phases, which quantify linear response characteristics, are given as follows:
\begin{align}
\label{eq:Z_s}
\boldsymbol{Z}_\mathrm{s}(x - \Phi, \Theta)
&= - \boldsymbol{U}_\mathrm{s}^*(x - \Phi, \Theta), \\
\label{eq:Z_t}
\boldsymbol{Z}_\mathrm{t}(x - \Phi, \Theta)
&= + \boldsymbol{U}_\mathrm{t}^*(x - \Phi, \Theta).
\end{align}
The phase equations, Eqs.~(\ref{eq:phase_equation_pertubed_Phi_rewrite}) and (\ref{eq:phase_equation_pertubed_Theta_rewrite}), indicate that the spatial and temporal phases are mutually coupled, resulting in nontrivial spatiotemporal phase dynamics.

\subsection{Traveling breathers with weak couplings}

\par
In this subsection, we consider weakly coupled reaction--diffusion systems, which are described by the following equation:
\begin{align}
\label{eq:coupled_RDsystems}
& \frac{\partial}{\partial t} \boldsymbol{X}_\sigma(x, t) \nonumber \\
& =  \boldsymbol{F}(\boldsymbol{X}_\sigma(x,t))
+ {\rm D} \frac{\partial^2}{\partial x^2} \boldsymbol{X}_\sigma
+ \varepsilon \boldsymbol{G}(\boldsymbol{X}_\sigma(x,t), \boldsymbol{X}_\tau(x,t)),
\end{align}
where subscripts $\sigma$ and $\tau$ denote indices of the systems
($(\sigma, \tau) = (1,2)$ or $(2,1)$), and the term
$\varepsilon \boldsymbol{G}(\boldsymbol{X}_\sigma(x,t), \boldsymbol{X}_\tau(x,t)) \in \mathbb{R}^N$ represents the coupling between the two systems at a spatial point $x$.
We assume a sufficiently weak coupling intensity ($\varepsilon \ll 1$) to ensure that a solution $\boldsymbol{X}_\sigma(x,t)$ remains in the vicinity of the orbit of the limit-torus solution $\boldsymbol{X}_0(x-\Phi, \Theta)$.
Under this assumption, the phase equations are obtained as follows:
\begin{align}
\label{eq:coupled_phase_equation_Phi}
& \dot{\Phi}_\sigma(t)
= c + \varepsilon \tilde{\Gamma}_\mathrm{s} \left(\Phi_\sigma - \Phi_\tau, \Theta_\sigma, \Theta_\tau \right), \\
\label{eq:coupled_phase_equation_Theta}
& \dot{\Theta}_\sigma(t)
= \omega + \varepsilon \tilde{\Gamma}_\mathrm{t} \left( \Phi_\sigma - \Phi_\tau, \Theta_\sigma, \Theta_\tau \right),
\end{align}
where 
\begin{align}
\label{eq:phase_coupling_function_non-averaged_phi}
\tilde{\Gamma}_\mathrm{s}&(\Phi_\sigma-\Phi_\tau,\Theta_\sigma,\Theta_\tau) \nonumber \\
= & \int_0^{2L}\mathrm{d}x\, \boldsymbol{Z}_\mathrm{s}(x-\Phi_\sigma,\Theta_\sigma) \nonumber \\
& \cdot \boldsymbol{G}( \boldsymbol{X}_0(x-\Phi_\sigma,\Theta_\sigma), \boldsymbol{X}_0(x-\Phi_\tau,\Theta_\tau) ), \\
\label{eq:phase_coupling_function_non-averaged_theta}
\tilde{\Gamma}_\mathrm{t}&(\Phi_\sigma-\Phi_\tau,\Theta_\sigma,\Theta_\tau) \nonumber \\
=& \int_0^{2L}\mathrm{d}x\, \boldsymbol{Z}_\mathrm{t}(x-\Phi_\sigma,\Theta_\sigma) \nonumber \\
& \cdot \boldsymbol{G}( \boldsymbol{X}_0(x-\Phi_\sigma,\Theta_\sigma), \boldsymbol{X}_0(x-\Phi_\tau,\Theta_\tau) ).
\end{align}
As seen in Eqs.~(\ref{eq:coupled_phase_equation_Phi})--(\ref{eq:phase_coupling_function_non-averaged_theta}), the phase coupling functions, $\tilde{\Gamma}_\mathrm{s}(\Phi_\sigma-\Phi_\tau,\Theta_\sigma,\Theta_\tau)$ and $\tilde{\Gamma}_\mathrm{t}(\Phi_\sigma-\Phi_\tau,\Theta_\sigma,\Theta_\tau)$, depend on the spatial phase difference between the systems and the temporal phases of both systems.

\par 
Below we introduce the slow phase variables as  
\begin{align}
\label{eq:slow_variable_phi}
\phi_\sigma(t)
&= \Phi_\sigma(t) - ct, \\
\label{eq:slow_variable_theta}
\theta_\sigma(t)
&= \Theta_\sigma(t) - \omega t,
\end{align}
Then, Eqs.~(\ref{eq:coupled_phase_equation_Phi}) and (\ref{eq:coupled_phase_equation_Theta}) can be rewritten as follows:
\begin{align}
\label{eq:slow_phase_equation_phi}
\dot{\phi}_\sigma(t)
&=\varepsilon\tilde{\Gamma}_\mathrm{s}(\phi_\sigma-\phi_\tau,\omega t+\theta_\sigma,\omega t+\theta_\tau), \\
\label{eq:slow_phase_equation_theta}
\dot{\theta}_\sigma(t)
&=\varepsilon\tilde{\Gamma}_\mathrm{t}(\phi_\sigma-\phi_\tau,\omega t+\theta_\sigma,\omega t+\theta_\tau).
\end{align}
Applying the averaging method with respect to the temporal phases, we obtain the following equations:
\begin{align}
\label{eq:slow_phase_equation_averaged_phi}
\dot{\phi}_\sigma(t)
&=\varepsilon\Gamma_\mathrm{s}(\phi_\sigma-\phi_\tau,\theta_\sigma-\theta_\tau), \\
\label{eq:slow_phase_equation_averaged_theta}
\dot{\theta}_\sigma(t)
&=\varepsilon\Gamma_\mathrm{t}(\phi_\sigma-\phi_\tau,\theta_\sigma-\theta_\tau),
\end{align}
where the phase coupling functions for the spatial and temporal phases are given by
\begin{align}
\label{eq:phase_coupling_function_averaged_phi}
& \Gamma_\mathrm{s}(\phi_\sigma-\phi_\tau,\theta_\sigma-\theta_\tau) \nonumber \\
& =\frac{1}{2\pi}\int_0^{2\pi}\mathrm{d}\lambda\, \tilde{\Gamma}_\mathrm{s}(\phi_\sigma-\phi_\tau,\lambda+\theta_\sigma,\lambda+\theta_\tau), \\
\label{eq:phase_coupling_function_averaged_theta}
& \Gamma_\mathrm{t}(\phi_\sigma-\phi_\tau,\theta_\sigma-\theta_\tau) \nonumber \\
& =\frac{1}{2\pi}\int_0^{2\pi}\mathrm{d}\lambda\, \tilde{\Gamma}_\mathrm{t}(\phi_\sigma-\phi_\tau,\lambda+\theta_\sigma,\lambda+\theta_\tau).
\end{align}
Therefore, we obtain the following phase equations:
\begin{align}
\label{eq:phase_equation_phi(last)} 
\dot{\Phi}_\sigma(t)
&= c + \varepsilon \Gamma_\mathrm{s}(\Phi_\sigma - \Phi_\tau, \Theta_\sigma - \Theta_\tau), \\
\label{eq:phase_equation_theta(last)} 
\dot{\Theta}_\sigma(t)
&= \omega + \varepsilon \Gamma_\mathrm{t}(\Phi_\sigma - \Phi_\tau, \Theta_\sigma - \Theta_\tau),
\end{align}
where the phase coupling functions are explicitly described as
\begin{align}
\label{eq:Gamma_s(last)}
\Gamma_\mathrm{s}&(\Phi_\sigma - \Phi_\tau, \Theta_\sigma - \Theta_\tau) \nonumber \\
= & \frac{1}{2\pi} \int_0^{2\pi} \mathrm{d}\lambda\, \int_0^{2L} \mathrm{d}x\, 
\boldsymbol{Z}_\mathrm{s}(x - \Phi_\sigma, \lambda + \Theta_\sigma) \nonumber \\
& \cdot \boldsymbol{G}(\boldsymbol{X}_0(x - \Phi_\sigma, \lambda + \Theta_\sigma), \boldsymbol{X}_0(x - \Phi_\tau, \lambda + \Theta_\tau)), \\
\label{eq:Gamma_t(last)}
\Gamma_\mathrm{t}&(\Phi_\sigma - \Phi_\tau, \Theta_\sigma - \Theta_\tau) \nonumber \\
= & \frac{1}{2\pi} \int_0^{2\pi} \mathrm{d}\lambda\, \int_0^{2L} \mathrm{d}x\, 
\boldsymbol{Z}_\mathrm{t}(x - \Phi_\sigma, \lambda + \Theta_\sigma)  \nonumber \\
& \cdot \boldsymbol{G}(\boldsymbol{X}_0(x - \Phi_\sigma, \lambda + \Theta_\sigma), \boldsymbol{X}_0(x - \Phi_\tau, \lambda + \Theta_\tau)).
\end{align}
The phase coupling functions, $\Gamma_\mathrm{s}(\Phi_\sigma - \Phi_\tau, \Theta_\sigma - \Theta_\tau)$ and $\Gamma_\mathrm{t}(\Phi_\sigma - \Phi_\tau, \Theta_\sigma - \Theta_\tau)$, depend only on the spatial and temporal phase differences.

\par 
We define the spatial and temporal phase differences as follows:
\begin{align}
\label{eq:Delta_Phi}
\Delta \Phi(t) &:= \Phi_1(t) - \Phi_2(t), \\
\label{eq:Delta_Theta}
\Delta \Theta(t) &:= \Theta_1(t) - \Theta_2(t).
\end{align}
From Eqs.~(\ref{eq:phase_equation_phi(last)}) and (\ref{eq:phase_equation_theta(last)}),
we obtain the following equations that describe the time evolution of the spatial and temporal phase differences:
\begin{align}
\label{eq:d(Delta_Phi)/dt}
\frac{\mathrm{d}}{\mathrm{d}t} \Delta \Phi(t)
&= \varepsilon \Gamma_\mathrm{s}^{(\mathrm{a})}(\Delta \Phi, \Delta \Theta),  \\
\label{eq:d(Delta_Theta)/dt}
\frac{\mathrm{d}}{\mathrm{d}t} \Delta \Theta(t)
&= \varepsilon \Gamma_\mathrm{t}^{(\mathrm{a})}(\Delta \Phi, \Delta \Theta).
\end{align}
The antisymmetric components of the phase coupling functions  
are denoted by $\Gamma_\mathrm{s}^{(\mathrm{a})}(\Delta \Phi, \Delta \Theta)$ and $\Gamma_\mathrm{t}^{(\mathrm{a})}(\Delta \Phi, \Delta \Theta)$. 
They are given by
\begin{align}
\label{eq:Gamma_s^a}
\Gamma_\mathrm{s}^{(\mathrm{a})}(\Delta \Phi, \Delta \Theta)
&= \Gamma_\mathrm{s}(\Delta \Phi, \Delta \Theta) - \Gamma_\mathrm{s}(-\Delta \Phi, -\Delta \Theta), \\
\label{eq:Gamma_t^a}
\Gamma_\mathrm{t}^{(\mathrm{a})}(\Delta \Phi, \Delta \Theta)
&= \Gamma_\mathrm{t}(\Delta \Phi, \Delta \Theta) - \Gamma_\mathrm{t}(-\Delta \Phi, -\Delta \Theta).
\end{align}
These two functions satisfy the following properties:
\begin{align}
\label{eq:antisymmertry(point)_Gamma_s^a}
\Gamma_\mathrm{s}^{(\mathrm{a})}(-\Delta \Phi, -\Delta \Theta)
&= -\Gamma_\mathrm{s}^{(\mathrm{a})}(\Delta \Phi, \Delta \Theta), \\
\label{eq:antisymmertry(point)_Gamma_t^a}
\Gamma_\mathrm{t}^{(\mathrm{a})}(-\Delta \Phi, -\Delta \Theta)
&= -\Gamma_\mathrm{t}^{(\mathrm{a})}(\Delta \Phi, \Delta \Theta),
\end{align}
which represent antisymmetry with respect to the origin $(\Delta \Phi, \Delta \Theta) = (0, 0)$.
It can be readily seen from Eqs.~(\ref{eq:d(Delta_Phi)/dt})--(\ref{eq:Gamma_t^a}) that the system has a fixed point at $(\Delta \Phi, \Delta \Theta) = (0, 0)$.

\par
Notably, we can analyze the stability of the in-phase synchronized state, i.e., $(\Delta \Phi, \Delta \Theta) = (0, 0)$.
By linearizing Eqs.~(\ref{eq:d(Delta_Phi)/dt}) and (\ref{eq:d(Delta_Theta)/dt}) around $(\Delta \Phi, \Delta \Theta) = (0, 0)$, we obtain the following equations:
\begin{align}
\label{eq:linearization}
& \frac{\mathrm{d}}{\mathrm{d}t}
\begin{pmatrix}
\Delta \Phi \\
\Delta \Theta
\end{pmatrix} \nonumber \\
& = 2 \varepsilon
\left(
\begin{array}{cc}
\partial_{\Delta \Phi} \Gamma_\mathrm{s} &
\partial_{\Delta \Theta} \Gamma_\mathrm{s} \\
\partial_{\Delta \Phi} \Gamma_\mathrm{t} &
\partial_{\Delta \Theta} \Gamma_\mathrm{t} 
\end{array}
\right) \Bigg|_{{\substack{(\Delta \Phi, \Delta \Theta) \\ = (0, 0)}}}
\begin{pmatrix}
\Delta \Phi \\
\Delta \Theta
\end{pmatrix},
\end{align}
where $\Gamma_\mathrm{p}(\Delta \Phi, \Delta \Theta)$ is simply denoted as $\Gamma_\mathrm{p}$ for $\mathrm{p} \in \{\mathrm{s}, \mathrm{t}\}$.
This equation describes the stability of the fixed point $(\Delta \Phi, \Delta \Theta) = (0,0)$.
We consider the following coupling function as a typical case:
\begin{align}
\label{eq:G_linear_coupling}
\varepsilon \boldsymbol{G}(\boldsymbol{X}_\sigma,\boldsymbol{X}_\tau)
=\varepsilon \left[\boldsymbol{X}_{\tau}(x,t)-\boldsymbol{X}_\sigma(x,t)\right],
\end{align}
which implies that each component is linearly coupled to its counterpart with equal weight.
Under the linear coupling described in Eq.~(\ref{eq:G_linear_coupling}), the linearized equation~(\ref{eq:linearization}) can be reduced to
\begin{align}
\label{eq:linearization(simple)}
& \frac{\mathrm{d}}{\mathrm{d}t}
\begin{pmatrix}
\Delta \Phi \\
\Delta \Theta
\end{pmatrix}
= -2 \varepsilon
\left(
\begin{array}{cc}
1 & 0 \\
0 & 1
\end{array}
\right)
\left(
\begin{array}{c}
\Delta \Phi \\
\Delta \Theta
\end{array}
\right).
\end{align}
This indicates that the in-phase synchronization with respect to both spatial and temporal phases, i.e., $(\Delta \Phi, \Delta \Theta) = (0, 0)$, is linearly stable for systems with the coupling described in Eq.~(\ref{eq:G_linear_coupling}).
The orthonormalization condition (Eq.~(\ref{eq:orthonormalization})) is used in deriving Eq.~(\ref{eq:linearization(simple)}).
A similar stability analysis of the in-phase synchronized state for limit-cycle solutions can be found in Ref.~\cite{kawamura_optimizing_2017}.

\par
Finally, we note that the forms of Eqs.~(\ref{eq:phase_equation_phi(last)}) and (\ref{eq:phase_equation_theta(last)}) are the same as those of the phase equations derived for weakly coupled limit-torus oscillators, which are described by finite-dimensional dynamical systems~\cite{izhikevich_weakly_1999,demirt_phase_2010}. 
Therefore, as with ordinary limit-torus oscillators, coupled reaction--diffusion systems with limit-torus solutions can be reduced to a set of phase equations.
We also note that similar phase equations for oscillatory convection systems with spatial translational symmetry have been derived in previous studies~\cite{kawamura_phase_description_2015,kawamura_phase_2019}.

\section{PHASE REDUCTION ANALYSIS OF TRAVELING BREATHERS}
\label{sec:NumericalAnalysis}

\par
In this section, we perform phase reduction analysis of coupled reaction--diffusion systems.
Specifically, we analyze the FHN models exhibiting standing breathers (Sec.~\ref{subsec:coupled_FHN})~\cite{hagberg_pattern_1994} and the Gray--Scott models exhibiting traveling breathers (Sec.~\ref{subsec:coupled_Gray--Scott_Model})~\cite{yadome_chaotic_2011}.
Additionally, we perform direct numerical simulations of the coupled reaction--diffusion systems and compare the time evolution of the spatial and temporal phase differences obtained from the phase equations with those obtained from the simulations.

\par 
The details of the phase reduction analysis presented in this section can be found in Appendices~\ref{appendix:A}--\ref{appendix:C} as supplementary information.   
Appendix~\ref{appendix:A} describes the procedure for calculating the adjoint zero eigenfunctions.
Appendix~\ref{appendix:B} describes the method for calculating the traveling velocity $c \neq 0$, which is employed in Sec.~\ref{subsec:coupled_Gray--Scott_Model}.  
Appendix~\ref{appendix:C} describes how the spatial and temporal phases are determined from the spatiotemporal dynamics obtained from the direct numerical simulations.

\par
In the calculation of the limit-torus solutions and the solution of the adjoint equation, the system was discretized using spatial grid sizes of $\Delta x = 2L / 2^{10}$ and $\Delta x = 2L / 2^{13}$ in the FHN and Gray--Scott models, respectively.
Time integration was performed using the explicit Euler scheme with time steps of $\Delta t = T_{\Theta}/2^{16}$ and $\Delta t = T_{\Theta}/2^{20}$ in the FHN and Gray--Scott models, respectively.
The constant $T_{\Theta} := 2\pi / \omega$ denotes the period of the temporal phase.
The traveling velocity and oscillation frequency of the unperturbed solution, $\boldsymbol{X}(x,t) = \boldsymbol{X}_0(x - \Phi(t), \Theta(t))$, in the FHN model were found to be $c = 0$ and $\omega \simeq 5.59023 \times 10^{-2}$, respectively.  
In the Gray--Scott model, the corresponding values were $c \simeq 4.31310 \times 10^{-2}$ and $\omega \simeq 2.78994 \times 10^{-2}$.
These values of $c$ and $\omega$ were used to solve the adjoint equation, which was integrated over total durations of $100 T_{\Theta}$ and $2700 T_{\Theta}$, in the FHN and Gray--Scott models, respectively.

\par
In the direct numerical simulations, the system was discretized using spatial grid sizes of $\Delta x = 2L / 2^{10}$ and $\Delta x = 2L / 2^{12}$ in the FHN and Gray--Scott models, respectively.
Time integration was performed using the explicit Euler scheme with time steps of $\Delta t = 0.002$ and $\Delta t = 0.0008$ in the FHN and Gray--Scott models, respectively.

\subsection{Standing breathers in the FHN models}
\label{subsec:coupled_FHN}

\par 
In this subsection, we examine the FHN models exhibiting standing breathers, described by the following equations:
\begin{align}
\label{eq:FHNmodel}
& \boldsymbol{X}_\sigma(x, t)
= \begin{pmatrix}
u_\sigma \\
v_\sigma
\end{pmatrix}, 
~~
\boldsymbol{F}(\boldsymbol{X}_\sigma(x,t))
= \begin{pmatrix}
u_\sigma - u_\sigma^3 - v_\sigma \\
\mu (u_\sigma - a_1 v_\sigma - a_0)
\end{pmatrix}, \\
\label{eq:CouplingFunction_G_FHN}
& \varepsilon \boldsymbol{G}(\boldsymbol{X}_\sigma(x,t), \boldsymbol{X}_\tau(x,t)) 
= \varepsilon {\rm K} (\boldsymbol{X}_\tau(x,t) - \boldsymbol{X}_\sigma(x,t)),
\end{align} 
where $u_\sigma := u_\sigma(x, t)$ and $v_\sigma := v_\sigma(x, t)$, and the coefficient matrix ${\rm K}$ is diagonal.  
The system has a size of $2L = 120$, which is sufficiently large compared to that of the breather.
The diffusion coefficient is ${\rm D} = \mathrm{diag}(1.0, 2.5)$.
The remaining parameters are $\mu = 0.03$, $a_0 = -0.1$, and $a_1 = 2.0$.
The solution $\boldsymbol{X}_{\sigma}(x, t)$ exhibits standing breathers.
Figure~\ref{fig:fig2} shows the limit-torus solution $\boldsymbol{X}_0(x - \Phi, \Theta)$ with $\Phi = 0$.
In this study, the spatial region where the values of $u_0(x-\Phi, \Theta)$ and $v_0(x-\Phi, \Theta)$ significantly differ from those in the surrounding region is referred to as the {\it core region} of the breather.
In the absence of coupling ($\varepsilon = 0$), the traveling velocity of the standing breather is zero, i.e.,
\begin{align}
\label{eq:c=0}
c = 0.
\end{align}
Furthermore, the limit-torus solution satisfies the following spatial reflection symmetry (see Fig.~\ref{fig:fig2}):
\begin{align}
\label{eq:reflection_symmetry_X0_FHN}
\boldsymbol{X}_0(-(x - \Phi), \Theta) = \boldsymbol{X}_0(x - \Phi, \Theta).
\end{align}
The zero traveling velocity is expected from the spatial reflection symmetry of both the limit-torus solution and the governing equations, the latter of which is mentioned in Sec.~\ref{subsec:RD-system}.

\par
The Floquet zero eigenfunctions $\boldsymbol{U}_\mathrm{s}(x-\Phi, \Theta)$ and $\boldsymbol{U}_\mathrm{t}(x-\Phi, \Theta)$ are determined using Eqs.~(\ref{eq:choice_of_Us}) and (\ref{eq:choice_of_Ut}). 
Because of the reflection symmetry of the limit-torus solution, the Floquet zero eigenfunctions exhibit the following reflection properties (see Fig.~\ref{fig:fig3}):
\begin{align}
\label{eq:Us_symmetry(FHN)}
\boldsymbol{U}_\mathrm{s}(-(x - \Phi), \Theta)
&= -\boldsymbol{U}_\mathrm{s}(x - \Phi, \Theta), \\
\label{eq:Ut_symmetry(FHN)}
\boldsymbol{U}_\mathrm{t}(-(x - \Phi), \Theta)
&= +\boldsymbol{U}_\mathrm{t}(x - \Phi, \Theta).
\end{align}
Consequently, the phase sensitivity functions (adjoint zero eigenfunctions) exhibit the following reflection properties (see Fig.~\ref{fig:fig4}):
\begin{align}
\label{eq:Zs_symmetry(FHN)}
\boldsymbol{Z}_\mathrm{s}(-(x - \Phi), \Theta)
&= -\boldsymbol{Z}_\mathrm{s}(x - \Phi, \Theta), \\
\label{eq:Zt_symmetry(FHN)}
\boldsymbol{Z}_\mathrm{t}(-(x - \Phi), \Theta)
&= +\boldsymbol{Z}_\mathrm{t}(x - \Phi, \Theta).
\end{align}
The Floquet zero eigenfunctions and phase sensitivity functions are localized around the core region of the breather.
Regarding the $u_\sigma$ component, the corresponding components of the Floquet zero eigenfunctions and phase sensitivity functions are localized near the edges of the core region.
Although this localization appears only at the edges, we still consider it to be one manifestation of the localization around the core region.

\par
Because of the spatial reflection properties of the limit-torus solutions and the phase sensitivity functions, the antisymmetric components of the phase coupling functions exhibit additional spatial reflection properties.
Using the phase differences $\Delta \Phi$ and $\Delta \Theta$, the phase coupling functions can be rewritten as
\begin{align}
\label{eq:Gamma_s(rewrite)}
\Gamma_\mathrm{s}&(\Delta \Phi, \Delta \Theta) \nonumber \\
= & \frac{1}{2\pi} \int_0^{2\pi} \mathrm{d}\lambda\, \int_0^{2L} \mathrm{d}x\,
\boldsymbol{Z}_\mathrm{s}(x - \Delta \Phi, \lambda + \Delta \Theta) \nonumber \\
& \cdot {\rm K}(\boldsymbol{X}_0(x, \lambda) - \boldsymbol{X}_0(x - \Delta \Phi, \lambda + \Delta \Theta)), \\
\label{eq:Gamma_t(rewrite)}
\Gamma_\mathrm{t}&(\Delta \Phi, \Delta \Theta) \nonumber \\
= & \frac{1}{2\pi} \int_0^{2\pi} \mathrm{d}\lambda\, \int_0^{2L} \mathrm{d}x\,
\boldsymbol{Z}_\mathrm{t}(x - \Delta \Phi, \lambda + \Delta \Theta) \nonumber \\
& \cdot {\rm K}(\boldsymbol{X}_0(x, \lambda) - \boldsymbol{X}_0(x - \Delta \Phi, \lambda + \Delta \Theta)).
\end{align}
From Eqs.~(\ref{eq:reflection_symmetry_X0_FHN}), (\ref{eq:Zs_symmetry(FHN)}), and (\ref{eq:Zt_symmetry(FHN)}), the phase coupling functions satisfy the following reflection properties:
\begin{align}
\label{eq:Gamma_s_symmetry}
& \Gamma_\mathrm{s}(-\Delta \Phi, \Delta \Theta)= -\Gamma_\mathrm{s}(\Delta \Phi, \Delta \Theta), \\
\label{eq:Gamma_t_symmetry}
& \Gamma_\mathrm{t}(-\Delta \Phi, \Delta \Theta)
= +\Gamma_\mathrm{t}(\Delta \Phi, \Delta \Theta).
\end{align}
Accordingly, the antisymmetric components of the phase coupling functions satisfy similar reflection properties:
\begin{align}
\label{eq:Gamma_s^a_symmetry}
& \Gamma_\mathrm{s}^{(\mathrm{a})}(-\Delta \Phi, \Delta \Theta)
= -\Gamma_\mathrm{s}^{(\mathrm{a})}(\Delta \Phi, \Delta \Theta), \\
\label{eq:Gamma_t^a_symmetry}
& \Gamma_\mathrm{t}^{(\mathrm{a})}(-\Delta \Phi, \Delta \Theta)
= +\Gamma_\mathrm{t}^{(\mathrm{a})}(\Delta \Phi, \Delta \Theta).
\end{align}
Furthermore, by combining Eqs.~(\ref{eq:antisymmertry(point)_Gamma_s^a}), (\ref{eq:antisymmertry(point)_Gamma_t^a}), (\ref{eq:Gamma_s^a_symmetry}), and (\ref{eq:Gamma_t^a_symmetry}), we obtain the following additional properties:
\begin{align}
\label{eq:Gamma_s^a_symmetry2}
&\Gamma_\mathrm{s}^{(\mathrm{a})}(\Delta \Phi, -\Delta \Theta)
= +\Gamma_\mathrm{s}^{(\mathrm{a})}(\Delta \Phi, \Delta \Theta), \\
\label{eq:Gamma_t^a_symmetry2}
&\Gamma_\mathrm{t}^{(\mathrm{a})}(\Delta \Phi, -\Delta \Theta)
= -\Gamma_\mathrm{t}^{(\mathrm{a})}(\Delta \Phi, \Delta \Theta).
\end{align}
From Eq.~(\ref{eq:Gamma_s^a_symmetry}), the nullclines of $\Gamma_\mathrm{s}^{(\mathrm{a})}(\Delta \Phi, \Delta \Theta)$ are located at $\Delta \Phi = 0$ and $L$.
Similarly, from Eq.~(\ref{eq:Gamma_t^a_symmetry2}), the nullclines of $\Gamma_\mathrm{t}^{(\mathrm{a})}(\Delta \Phi, \Delta \Theta)$ are located at $\Delta \Theta = 0$ and $\pi$.
Therefore, the rectangular region defined by $0 \leq \Delta\Phi/L \leq 1$ and $0 \leq \Delta\Theta/\pi \leq 1$ is surrounded by four nullclines and contains four fixed points located at $(\Delta \Phi / L, \Delta \Theta / \pi) = (0,0)$, $(1,0)$, $(0,1)$, and $(1,1)$.  
Taking Eqs.~(\ref{eq:antisymmertry(point)_Gamma_s^a}), (\ref{eq:antisymmertry(point)_Gamma_t^a}), and (\ref{eq:Gamma_s^a_symmetry})--(\ref{eq:Gamma_t^a_symmetry2}) into consideration, the other three rectangular regions, which are symmetric to the specified one with respect to the origin and the axes, also exhibit similar configurations of nullclines and fixed points.

\par
Here, we consider two types of coupling for the FHN model:
(i) coupling via both $u_\sigma$ and $v_\sigma$, with ${\rm K} = \mathrm{diag}(1,1)$; 
(ii) coupling via $u_\sigma$ alone, with ${\rm K} = \mathrm{diag}(1,0)$.  
Although both types of coupling are examined in this study, it is more common in FHN models to consider coupling only via the membrane potential variable $u_\sigma$.

\par 
\textbf{Case (i).} 
The antisymmetric components of the phase coupling functions are shown in Fig.~\ref{fig:fig5}.  
They exhibit antisymmetry with respect to the origin (see Eqs.~(\ref{eq:antisymmertry(point)_Gamma_s^a}) and (\ref{eq:antisymmertry(point)_Gamma_t^a})).
Additionally, they satisfy the properties described in Eqs.~(\ref{eq:Gamma_s^a_symmetry})--(\ref{eq:Gamma_t^a_symmetry2}).
Figure~\ref{fig:fig6} presents comparisons between the time evolution of phase differences obtained from simulations of the FHN models with coupling intensity $\varepsilon$ and the corresponding theoretical predictions.
Figures~\ref{fig:fig6}(a) and \ref{fig:fig6}(b) show representative trajectories converging to the stable fixed points $(\Delta \Phi/L, \Delta \Theta/\pi) = (0,0)$ and $(1,1)$, respectively. 
Figure~\ref{fig:fig6}(c) shows these trajectories in the $(\Delta \Phi/L, \Delta \Theta/\pi)$ space along with nullclines.
The convergence to $(0,0)$ is observed only for a relatively narrow range of the initial phase differences, whereas the convergence to $(1,1)$ is observed for a much broader range.
The convergence to $(1,1)$ is significantly slower than that to $(0,0)$ because $|\Gamma^{(\mathrm{a})}_\mathrm{t}(\Delta \Phi, \Delta \Theta)|$ is small near $\Delta \Phi / L = 1$.
Note that the fixed point $(0,0)$ must be linearly stable because Eq.~(\ref{eq:linearization(simple)}) was derived under the condition ${\rm K} = \mathrm{diag}(1,1)$ with the coupling function given in Eq.~(\ref{eq:CouplingFunction_G_FHN}).

\par
\textbf{Case (ii).}
Figure~\ref{fig:fig7} shows the antisymmetric components of the phase coupling functions.
As in case~$\mathrm{(i)}$, both the antisymmetry with respect to the origin (see Eqs.~(\ref{eq:antisymmertry(point)_Gamma_s^a}) and (\ref{eq:antisymmertry(point)_Gamma_t^a})) and the properties described in Eqs.~(\ref{eq:Gamma_s^a_symmetry})--(\ref{eq:Gamma_t^a_symmetry2}) are satisfied.
As in Fig.~\ref{fig:fig6}, Fig.~\ref{fig:fig8} presents comparisons between the theoretical values and the results obtained from direct numerical simulations.
Figures~\ref{fig:fig8}(a) and \ref{fig:fig8}(b) show representative trajectories converging to the fixed points $(\Delta \Phi/L, \Delta \Theta/\pi) = (0,1)$ and $(1,0)$, respectively, both of which are linearly stable.
Figure~\ref{fig:fig8}(c) shows these trajectories in the $(\Delta \Phi/L, \Delta \Theta/\pi)$ space along with nullclines.
We observe that the convergence to $(0,1)$ and $(1,0)$ is approximately separated by the nullcline of $\Gamma_\mathrm{s}^{(\mathrm{a})}(\Delta \Phi, \Delta \Theta)$.
When the trajectory converges to $(0,1)$, $\Delta \Phi$ decreases monotonically, while $\Delta \Theta$ initially oscillates and then converges to the fixed point.
The trajectories converging to $(1,0)$ approach the fixed point monotonically for both $\Delta \Phi$ and $\Delta \Theta$ over a large part of the $(\Delta \Phi/L, \Delta \Theta/\pi)$ space.
The convergence to $(1,0)$ is significantly slower than that to $(0,1)$ because $|\Gamma^{(\mathrm{a})}_\mathrm{s}(\Delta \Phi, \Delta \Theta)|$ and $|\Gamma^{(\mathrm{a})}_\mathrm{t}(\Delta \Phi, \Delta \Theta)|$ are small near $\Delta \Phi / L = 1$.
It is worth noting that Eq.~(\ref{eq:linearization(simple)}) cannot be derived under the condition ${\rm K} = \mathrm{diag}(1,0)$ with the coupling function given in Eq.~(\ref{eq:CouplingFunction_G_FHN}). 
Therefore, it cannot be ensured that the fixed point $(0,0)$ is linearly stable. 
Indeed, Fig.~\ref{fig:fig8}(c) indicates that it is a saddle point.

\par
In both cases, the antisymmetric components of the phase coupling functions exhibit a characteristic structure centered at $\Delta \Phi / L = 0$ (see Figs.~\ref{fig:fig5} and \ref{fig:fig7}).  
The width of this {\it characteristic region}, measured in terms of $\Delta \Phi$, is independent of the system size.  
This independence is justified by the fact that the size of the breather is independent of the system size. 
Consequently, the Floquet zero eigenfunctions and phase sensitivity functions are independent of the system size because they are localized around the core region (see the fifth paragraph of Sec.~\ref{sec:concluding_remarks} for details).

\subsection{Traveling breathers in the Gray--Scott models}
\label{subsec:coupled_Gray--Scott_Model}

\par
In this subsection, we consider the Gray--Scott model exhibiting traveling breathers ($c \neq 0$), which is described by
\begin{align}
\label{eq:GSM_model}
& \boldsymbol{X}_\sigma(x, t)
= \begin{pmatrix}
u_\sigma \\
v_\sigma
\end{pmatrix}, 
~
\boldsymbol{F}(\boldsymbol{X}_\sigma(x,t))
=
\begin{pmatrix}
u_\sigma^2 v_\sigma - (f + k) u_\sigma \\
- u_\sigma^2 v_\sigma + f (1 - v_\sigma) 
\end{pmatrix}, \\
\label{eq:CouplingFunction_G_GSM}
& \varepsilon \boldsymbol{G}(\boldsymbol{X}_\sigma(x,t), \boldsymbol{X}_\tau(x,t)) 
= \varepsilon (\boldsymbol{X}_\tau(x,t) - \boldsymbol{X}_\sigma(x,t)),
\end{align}
where $u_\sigma := u_\sigma(x, t)$ and $v_\sigma := v_\sigma(x, t)$.
The system has a size of $2L = 250$, which is sufficiently large compared to that of the breather.
The diffusion coefficient is ${\rm D} = \mathrm{diag}(1.0, 1.9)$.
The other parameters are set to $f = 0.018$ and $k = 0.052$.
The time series $\boldsymbol{X}_1(x, t)$ in the absence of coupling ($\varepsilon = 0$) is presented in Fig.~\ref{fig:fig9}.  
We observe that the traveling breather exhibits the spatial and temporal phases $\Phi_1 = ct$ and $\Theta_1 = \omega t$, respectively.
Figure~\ref{fig:fig10} shows the limit-torus solution $\boldsymbol{X}_0(x - \Phi, \Theta)$ with $\Phi = 0$.
The spatiotemporal pattern of $\boldsymbol{X}_0(x - \Phi, \Theta)$ breaks the spatial reflection symmetry; this can be clearly observed in the spatial asymmetry of the core region (see Fig.~\ref{fig:fig10}).
Thus, the limit-torus solution does not satisfy Eq.~(\ref{eq:reflection_symmetry_X0_FHN}).

\par 
Figure~\ref{fig:fig11} shows the Floquet zero eigenfunctions $\boldsymbol{U}_\mathrm{s}(x - \Phi, \Theta)$ and $\boldsymbol{U}_\mathrm{t}(x - \Phi, \Theta)$. Figure~\ref{fig:fig12} shows the corresponding phase sensitivity functions $\boldsymbol{Z}_\mathrm{s}(x - \Phi, \Theta)$ and $\boldsymbol{Z}_\mathrm{t}(x - \Phi, \Theta)$.
As in the FHN model, the Floquet zero eigenfunctions and phase sensitivity functions are localized around the core region of the breather. 
Since the size of the breather is independent of the system size, the Floquet zero eigenfunctions and phase sensitivity functions are also independent of the system size.
The antisymmetric components of the phase coupling functions, $\Gamma_\mathrm{s}^{(\mathrm{a})}(\Delta \Phi, \Delta \Theta)$ and $\Gamma_\mathrm{t}^{(\mathrm{a})}(\Delta \Phi, \Delta \Theta)$, are shown in Fig.~\ref{fig:fig13}.
They exhibit antisymmetry with respect to the origin (see Eqs.~(\ref{eq:antisymmertry(point)_Gamma_s^a}) and (\ref{eq:antisymmertry(point)_Gamma_t^a})). 
Since Eq.~(\ref{eq:reflection_symmetry_X0_FHN}) does not hold, the properties described in Eqs.~(\ref{eq:Us_symmetry(FHN)})--(\ref{eq:Zt_symmetry(FHN)}) and (\ref{eq:Gamma_s_symmetry})--(\ref{eq:Gamma_t^a_symmetry2}) are not satisfied.
Note that the origin must be a linearly stable fixed point because Eq.~(\ref{eq:linearization(simple)}) was derived under the coupling function described in Eq.~(\ref{eq:CouplingFunction_G_GSM}).

\par 
As in Figs.~\ref{fig:fig6} and \ref{fig:fig8}, Fig.~\ref{fig:fig14} shows comparisons between the theoretical values and the results obtained from direct numerical simulations.
Figures~\ref{fig:fig14}(a)--\ref{fig:fig14}(c) show representative trajectories converging to the stable fixed point $(\Delta \Phi/L, \Delta \Theta/\pi) = (0,0)$. Figures~\ref{fig:fig14}(d)--\ref{fig:fig14}(f) show these trajectories in the $(\Delta \Phi/L, \Delta \Theta/\pi)$ space along with the nullclines.
These figures show that the spatial and temporal in-phase state is globally stable. 
In addition, its linear stability is ensured by Eq.~(\ref{eq:linearization(simple)}).
Figures~\ref{fig:fig14}(a) and \ref{fig:fig14}(d) show that the temporal and spatial phase differences evolve monotonically in most parts of the space.
Both phase differences exhibit characteristic behavior in the region of $|\Delta \Phi / L| \lesssim 0.5$. 
They evolve rapidly throughout this region, except near $\Delta \Phi / L \simeq 0$, where the velocities of the spatial and temporal phase differences decrease significantly.
We refer to this region as the {\it characteristic region}, following the terminology used in Sec.~\ref{subsec:coupled_FHN}. 
The width of the characteristic region in the antisymmetric components of the phase coupling functions, measured in terms of $\Delta \Phi$, is independent of the system size. 
As mentioned in Sec.~\ref{subsec:coupled_FHN}, this independence is justified by the fact that the size of the breather is independent of the system size. 
Consequently, the Floquet zero eigenfunctions and phase sensitivity functions are independent of the system size (see the fifth paragraph of Sec.~\ref{sec:concluding_remarks} for details).

\par
The time evolution of the phase differences at the center of the characteristic region exhibits nontrivial behavior.
For example, Figs.~\ref{fig:fig14}(a) and \ref{fig:fig14}(d) show nearly monotonic convergence with a slight curvature near the origin $(\Delta \Phi/L, \Delta \Theta/\pi) = (0, 0)$.
In contrast, Figs.~\ref{fig:fig14}(b) and \ref{fig:fig14}(e) display a large swing of $\Delta \Theta$ during the convergence to the fixed point.
Finally, Figs.~\ref{fig:fig14}(c) and \ref{fig:fig14}(f) show a representative trajectory starting near the unstable fixed point.  
Notably, while $\Delta \Theta$ decreases monotonically, $\Delta \Phi$ becomes negative during the  convergence, despite starting from a positive value.
The transition of $\Delta \Phi$ from a positive to a negative value highlights the nontrivial nature of phase dynamics.
Furthermore, the insets in Figs.~\ref{fig:fig14}(c) and \ref{fig:fig14}(f) show that the trajectory intersects the nullcline of $\Gamma_\mathrm{s}^{(\mathrm{a})}(\Delta \Phi, \Delta \Theta)$ within the time range $\varepsilon t < 1$.
This early-time behavior, where $\Delta \Phi$ initially increases and then decreases, reflects the configurations of the nullclines.

\section{CONCLUDING REMARKS}
\label{sec:concluding_remarks}

\par
In this paper, we formulated a theory for the phase reduction analysis of reaction--diffusion systems exhibiting traveling and oscillating spatiotemporal patterns, namely, limit-torus solutions to partial differential equations with spatial translational symmetry~(Sec.~\ref{sec:Theory}). 
Then, we performed phase reduction analysis of the FHN models exhibiting standing breathers and Gray--Scott models exhibiting traveling breathers~(Sec.~\ref{sec:NumericalAnalysis}).
Figures~\ref{fig:fig6}, \ref{fig:fig8}, and~\ref{fig:fig14} show our main results. 
We observe that the phase reduction analysis successfully captures the nontrivial spatiotemporal phase dynamics. 
The theoretical values obtained using phase reduction analysis are consistent with the results obtained from the direct numerical simulations.

\par
For a pair of FHN models with ${\rm K} = \mathrm{diag}(1,1)$, using Eq.~(\ref{eq:linearization(simple)}), we found that the fixed point corresponding to the in-phase synchronized state with respect to both the spatial and temporal phases is linearly stable.
However, this synchronized state is reached only from a relatively narrow range of initial conditions for the phase differences (Fig.~\ref{fig:fig6}).  
For FHN models with ${\rm K} = \mathrm{diag}(1,0)$, the system converges to one of the two distinct states: $\mathrm{(i)}$ in-phase synchronization with respect to the spatial phase and anti-phase synchronization with respect to the temporal phase, or $\mathrm{(ii)}$ synchronization with the opposite configuration (Fig.~\ref{fig:fig8}). 
During convergence to the former, the temporal phase difference exhibits transient oscillations before settling into a steady state, whereas the convergence to the latter proceeds monotonically over most of the $(\Delta \Phi, \Delta \Theta)$ space (Fig.~\ref{fig:fig8}(c)). 
Figures~\ref{fig:fig6}(b) and \ref{fig:fig8}(b) show that convergence to the anti-phase state with respect to the spatial phase is slow.
The underlying reason is discussed in the fourth paragraph of this section.

\par
Furthermore, we performed phase reduction analysis of a pair of Gray--Scott models. 
The results showed that the in-phase synchronized state with respect to both the spatial and temporal phases is globally stable. 
Although the models exhibit global stability, the dynamics of the phase differences near $\Delta \Phi/L = 0$ exhibit nontrivial behavior, ranging from a nearly monotonic convergence to large transient swings (Figs.~\ref{fig:fig14}(d) and \ref{fig:fig14}(e)).  
Notably, during convergence to the in-phase synchronized state at $(\Delta \Phi/L, \Delta \Theta / \pi) = (0, 0)$, the spatial phase difference becomes negative, despite starting from a positive value (Fig.~\ref{fig:fig14}(f)).
This result indicates that during the convergence process, the relative positions of the breathers can be reversed.
We emphasize that the phase reduction analysis successfully captured the time evolution of both the spatial and temporal phase differences observed in direct numerical simulations, even in the case of traveling breathers.

\par 
In this study, we investigated standing and traveling breathers and found that their localized structure significantly affects their spatiotemporal phase dynamics.
In particular, for the FHN models, slow convergence toward the anti-phase synchronization with respect to the spatial phase was observed (Figs.~\ref{fig:fig6}(b) and \ref{fig:fig8}(b)).
This is attributed to the localization of the phase sensitivity functions around the core region of the breather (Fig.~\ref{fig:fig4}).
Specifically, when the coupling function is $\boldsymbol{G}(\boldsymbol{X}_\sigma(x,t), \boldsymbol{X}_\tau(x,t)) = {\rm K} (\boldsymbol{X}_\tau(x,t) - \boldsymbol{X}_\sigma(x,t))$, 
the antisymmetric component of the phase coupling function can be written as 
\begin{align}
\label{eq:Gamma_p^a_in_discussion}
\Gamma_\mathrm{p}^{(\mathrm{a})} & (\Delta \Phi, \Delta \Theta) 
= \frac{1}{2\pi}\int_0^{2\pi} \mathrm{d}\lambda\, \int_0^{2L} \mathrm{d}x\,  \boldsymbol{Z}_\mathrm{p}(x, \lambda) \nonumber \\
& \cdot {\rm K} [\boldsymbol{X}_0(x+\Delta \Phi, \lambda-\Delta \Theta)-\boldsymbol{X}_0(x-\Delta \Phi, \lambda+\Delta \Theta)],
\end{align}
where $\mathrm{p} \in \{\mathrm{s}, \mathrm{t}\}$.
For $\Delta \Phi/L \simeq 1$, the regions where 
$|\boldsymbol{Z}_\mathrm{p}(x, \lambda)|$ is localized do not overlap with those where the difference 
$|\boldsymbol{X}_0(x+\Delta \Phi, \lambda-\Delta \Theta) - \boldsymbol{X}_0(x-\Delta \Phi, \lambda+\Delta \Theta)|$ is large, resulting in a small magnitude of the integrand.
Consequently, the magnitudes of the antisymmetric components of the phase coupling functions, $|\Gamma_\mathrm{s}^{(\mathrm{a})}(\Delta \Phi, \Delta \Theta)|$ and $|\Gamma_\mathrm{t}^{(\mathrm{a})}(\Delta \Phi, \Delta \Theta)|$, decrease when $\Delta \Phi / L \simeq 1$ (Figs.~\ref{fig:fig5} and \ref{fig:fig7}).
A similar tendency was observed in the case of the coupled Gray--Scott models, where ${\rm K} = \mathrm{diag}(1, 1)$ was employed (for simplicity, ${\rm K}$ was omitted in Eqs.~(\ref{eq:GSM_model}) and (\ref{eq:CouplingFunction_G_GSM})).
As shown in Fig.~\ref{fig:fig12}, the phase sensitivity functions are spatially localized around the core region of the breather, which leads to a decrease in the magnitudes of the antisymmetric components of the phase coupling functions near $\Delta \Phi/L = 1$ (Fig.~\ref{fig:fig13}).

\par 
Another important finding is that the Floquet zero eigenfunctions and phase sensitivity functions are localized around the core region, reflecting the localized structures of both traveling and standing breathers.
Because the size of the breather is independent of the system size, the Floquet zero eigenfunctions and phase sensitivity functions are also independent of the system size.
On the basis of Eq.~(\ref{eq:Gamma_p^a_in_discussion}) and the localized nature of the breather and phase sensitivity functions, the width of the characteristic region in the antisymmetric components of the phase coupling functions does not depend on the system size.
In this study, the region centered at $\Delta\Phi/L = 0$, where $\Gamma_\mathrm{s}^{(\mathrm{a})}(\Delta \Phi, \Delta \Theta)$ and $\Gamma_\mathrm{t}^{(\mathrm{a})}(\Delta \Phi, \Delta \Theta)$ exhibit distinct structures, is referred to as the characteristic region.
As the system size increases, only the region outside the characteristic region expands, whereas the width of the characteristic region remains unchanged. 
The magnitudes of the antisymmetric components of the phase coupling functions are small outside this region.
It is worth noting that the structure of the characteristic region, which reveals nontrivial behavior during convergence to the in-phase synchronized state with respect to the spatial phase, reflects the detailed structure of the limit-torus solutions and the corresponding phase sensitivity functions.

\par 
Although previous studies~\cite{kawamura_phase_2019,kawamura_phase_description_2015} performed phase reduction analysis only on systems without spatial reflection symmetry breaking, the present study demonstrates that the same framework is applicable to systems both with and without such symmetry breaking.
The limit-torus solution of a traveling breather breaks spatial reflection symmetry (Fig.~\ref{fig:fig10}). 
Accordingly, the Floquet zero eigenfunctions and the phase sensitivity functions exhibit spatial asymmetry that reflects the structure of the breather (Figs.~\ref{fig:fig11} and \ref{fig:fig12}).
Regardless of whether spatial reflection symmetry is broken or not, the phase reduction analysis can successfully calculate the Floquet zero eigenfunctions and phase sensitivity functions.
Considering its generality, the theory formulated in this study can be used in a wide range of applications, including systems with different model parameters and various physical settings.
For example, the theory formulated in this study can be extended to reaction--diffusion systems with chemotaxis~\cite{kawaguchi_pulse_2022}, as well as to classic models such as the Oregonator~\cite{dockery_numerical_1998,berenstein_cross-diffusion_2013,nicola_wave_2006}.
Analyzing a wide range of models can provide insights into the effect of the spatial distance between spatiotemporal patterns on their synchronization.

\par
The importance of phase reduction theory for traveling and oscillating spatiotemporal patterns is not limited to reaction--diffusion systems.  
Phase reduction frameworks for limit-torus solutions have been developed for the analysis of fluid dynamics~\cite{kawamura_phase_description_2015,kawamura_phase_2019}.  
Building upon these frameworks, future studies are expected to explore their application to rotating and oscillating convection (i.e., amplitude vacillation~\cite{lappa_rotating_2012,ghil_geophysical_2010,read_quasi-periodic_1992}) in systems of rotating fluid annuli~\cite{read_phase_2017,castrejon-pita_synchronization_2010,eccles_synchronization_2009,oshima_synchronization_2025}.
These systems, which have been experimentally investigated, can be regarded as analogs to atmospheric circulation systems.
Therefore, phase reduction analysis of the traveling and oscillating spatiotemporal patterns that emerge in rotating fluid annuli can provide novel insights into meteorological phenomena, such as synchronized atmospheric blocking events between the Northern and Southern hemispheres~\cite{duane_co-occurrence_1999}.

\par 
The development of the theory for phase reduction analysis is essential for understanding the synchronization properties inherent in the governing equations and contributes to the development of data-driven approaches.
Data-driven approaches based on the theory for phase reduction analysis can reveal the synchronization properties of various phenomena without requiring detailed assumptions or prior knowledge of the underlying governing equations~\cite{kralemann_vivo_2013,ota_direct_2014,stankovski_inference_2012,stankovski_coupling_2017,pikovsky_synchronization_2001,arai_extracting_2022,arai_interlimb_2024,funato_evaluation_2016,onojima_dynamical_2018,ota_interaction_2020}.
Although several data-driven methods for identifying the spatiotemporal phase dynamics of perturbed limit-cycle solutions have been recently proposed ~\cite{arai_setting_2025,fukami_data-driven_2024,yawata_phase_2025}, no such method has yet been proposed for limit-torus solutions. 
Our study, along with Refs.~\cite{kawamura_phase_2019,kawamura_phase_description_2015}, provides a foundation for the further development of data-driven approaches to investigate the spatiotemporal phase dynamics of traveling and oscillating patterns.
The method described in Appendix~\ref{appendix:C}, which successfully determines the spatial and temporal phases from the spatiotemporal dynamics obtained from direct numerical simulations, can also underpin such developments.


\begin{acknowledgments}
This work was supported by JSPS KAKENHI Grant Numbers JP24K23908, JP24K06910, JP25K01160.
Numerical simulations were conducted using Earth Simulator at JAMSTEC.
\end{acknowledgments}

\appendix

\begin{widetext}
\vspace{0.5em}  
\section{CALCULATION OF THE ADJOINT ZERO EIGENFUNCTIONS}
\label{appendix:A}

\par
In this appendix, we describe the procedure for calculating the adjoint zero eigenfunctions.  
In the phase reduction analysis of the FHN model and Gray--Scott model (see Sec.~\ref{sec:NumericalAnalysis}), we perform the following steps:
\begin{enumerate}
    \item Calculate frequency $\omega$ using the following time series:  
    \begin{align}
    \label{eq:X_bar_in_step1}
    \overline{\boldsymbol{X}}(\Theta(t)) 
    & = \int_0^{2L} \mathrm{d}x\, \boldsymbol{X}(x,t)
    \nonumber \\
    & = \int_0^{2L} \mathrm{d}x\, \boldsymbol{X}_0(x - \Phi(t), \Theta(t) ),
    \end{align}
    where $\boldsymbol{X}(x,t) = \boldsymbol{X}_0(x - \Phi(t), \Theta(t))$ is obtained by solving Eq.~(\ref{eq:RD-systems}).
    The time series $\overline{\boldsymbol{X}}(\Theta(t))$ depends only on the temporal phase $\Theta(t)$ since the integration over the spatial domain of size $2L$ eliminates the dependence on the spatial phase $\Phi(t)$.

    \item If $c \neq 0$, determine the traveling velocity $c$ using the time series $\boldsymbol{X}(x,t) = \boldsymbol{X}_0(x - \Phi(t), \Theta(t))$, which is obtained by solving Eq.~(\ref{eq:RD-systems}) (see Appendix~\ref{appendix:B}).

    \item If $c \neq 0$, update $\omega$ using the time series $\overline{\boldsymbol{X}}(\Theta(t))$, which is obtained by solving the following equation:
    \begin{align}
    \label{eq:RD-systems_with_gradient}
    \frac{\partial}{\partial t} \boldsymbol{X}(\chi, t)
    = \boldsymbol{F}(\boldsymbol{X}(\chi, t))
    + {\rm D} \frac{\partial^2}{\partial \chi^2} \boldsymbol{X}
    + c \frac{\partial}{\partial \chi} \boldsymbol{X}.
    \end{align}
    The above equation is obtained by introducing the comoving coordinate $\chi := x - ct$ into Eq.~(\ref{eq:RD-systems}). 
    When the traveling velocity is zero ($c = 0$), Eqs.~(\ref{eq:RD-systems}) and (\ref{eq:RD-systems_with_gradient}) coincide. 
    This step is performed to correct for the slight discrepancy in $\omega$ that arises from numerical errors in the time integration of Eqs.~(\ref{eq:RD-systems}) and (\ref{eq:RD-systems_with_gradient}).

    \item Compute the limit-torus solution $\boldsymbol{X}_0(x - \Phi, \Theta)$ with $\Phi = 0$ and $\Theta \in [0, 2\pi)$ by solving Eq.~(\ref{eq:RD-systems_with_gradient}) over one period of the temporal phase, $T_\Theta := 2\pi / \omega$.
    The state corresponding to $\Phi = 0$ satisfies the condition described in Eq.~(\ref{eq:value_H}) for $c = 0$ and the conditions described in Eqs.~(\ref{eq:condition_funcB}) and (\ref{eq:function_B}) for $c \neq 0$.
    (See Appendix~\ref{appendix:C} for the details of these equations.)

    \item Set the initial condition of $\tilde{\boldsymbol{U}}_\mathrm{p}^*(x-\Phi, 2\pi)$ with $\Phi = 0$ to the corresponding Floquet zero eigenfunction $\boldsymbol{U}_\mathrm{p}(x-\Phi, 2\pi)$ with $\Phi = 0$ ($\mathrm{p} \in \{\mathrm{s}, \mathrm{t}\})$. 
    Here, $\tilde{\boldsymbol{U}}_\mathrm{p}^*(x - \Phi, 2\pi)$ denotes the solution to the adjoint equation~(\ref{eq:adjoint_equation_2}), which is obtained by time integration (see step 6-{\rm A}).

    \item Iterate steps 6-{\rm A}–6-{\rm C} described below until the solution to the adjoint equation~(\ref{eq:adjoint_equation_2}) converges.
    Throughout these steps, we set $\Phi = 0$.
    
    \begin{enumerate}
        \item[6-{\rm A}.] Update $\tilde{\boldsymbol{U}}_\mathrm{s}^*(x-\Phi, \Theta)$ and $\tilde{\boldsymbol{U}}_\mathrm{t}^*(x-\Phi, \Theta)$ by solving Eq.~(\ref{eq:adjoint_equation_2}) over $T_\Theta$.

        \item[6-{\rm B}.] Perform orthonormalization to obtain $\boldsymbol{U}_\mathrm{s}^*(x-\Phi, \Theta)$ and $\boldsymbol{U}_\mathrm{t}^*(x-\Phi, \Theta)$ (see Eqs.~(\ref{eq:orthonormalization_s}) and (\ref{eq:orthonormalization_t})).

        \item[6-{\rm C}.] Replace $\tilde{\boldsymbol{U}}_\mathrm{s}^*(x-\Phi, \Theta)$ and $\tilde{\boldsymbol{U}}_\mathrm{t}^*(x-\Phi, \Theta)$ with $\boldsymbol{U}_\mathrm{s}^*(x-\Phi, \Theta)$ and $\boldsymbol{U}_\mathrm{t}^*(x-\Phi, \Theta)$, respectively.
    \end{enumerate}
\end{enumerate}

\par
Any linear combination of the two adjoint zero eigenfunctions, each associated with a zero eigenvalue and satisfying Eq.~(\ref{eq:adjoint_equation_2}), still constitutes a solution to the same equation.
Therefore, orthonormalization between the Floquet zero eigenfunctions and the corresponding adjoint zero eigenfunctions is required to uniquely determine $\boldsymbol{U}_\mathrm{s}^*(x - \Phi, \Theta)$ and $\boldsymbol{U}_\mathrm{t}^*(x - \Phi, \Theta)$.
In step~6-{\rm B}, we calculate functions $\boldsymbol{U}_\mathrm{s}^*(x-\Phi, \Theta)$ and $\boldsymbol{U}_\mathrm{t}^*(x-\Phi, \Theta)$ as follows:
\begin{align}
\label{eq:orthonormalization_s}
\boldsymbol{U}_\mathrm{s}^*(x-\Phi, \Theta)
&= c \bigl( \tilde{\boldsymbol{U}}_\mathrm{s}^*(x-\Phi, \Theta) - a \tilde{\boldsymbol{U}}_\mathrm{t}^*(x-\Phi, \Theta) \bigr), \\
\label{eq:orthonormalization_t}
\boldsymbol{U}_\mathrm{t}^*(x-\Phi, \Theta) 
&= d \bigl( \tilde{\boldsymbol{U}}_\mathrm{t}^*(x-\Phi, \Theta) - b \tilde{\boldsymbol{U}}_\mathrm{s}^*(x-\Phi, \Theta) \bigr).
\end{align}
To perform orthonormalization, these functions are constructed so as to satisfy the following condition:
\begin{align}
\label{eq:condition_for_tilde_Up}
\left\llbracket \boldsymbol{U}_\mathrm{p}^*(x-\Phi, \Theta), \boldsymbol{U}_\mathrm{q}(x-\Phi, \Theta) \right\rrbracket = \delta_{\mathrm{pq}}
~~
(\mathrm{p}, \mathrm{q} \in \{ \mathrm{s}, \mathrm{t} \}),
\end{align}
where $\delta_{\mathrm{pq}}$ denotes the Kronecker delta.
The condition described in Eq.~(\ref{eq:condition_for_tilde_Up}) is satisfied by selecting $a$, $b$, $c$, and $d$ as follows:
\begin{align}
a &= \frac{\sigma_{\mathrm{st}}}{\sigma_{\mathrm{tt}}}, &
b &= \frac{\sigma_{\mathrm{ts}}}{\sigma_{\mathrm{ss}}}, \\
c &= \frac{\sigma_{\mathrm{tt}}}{\sigma_{\mathrm{ss}} \sigma_{\mathrm{tt}} - \sigma_{\mathrm{st}} \sigma_{\mathrm{ts}}}, &
d &= \frac{\sigma_{\mathrm{ss}}}{\sigma_{\mathrm{ss}} \sigma_{\mathrm{tt}} - \sigma_{\mathrm{st}} \sigma_{\mathrm{ts}}}.
\end{align}
The inner product $\sigma_{\mathrm{pq}}$ is defined as:
\begin{align}
\sigma_{\mathrm{pq}} := \left\llbracket \tilde{\boldsymbol{U}}_\mathrm{p}^*(x-\Phi, \Theta), \boldsymbol{U}_\mathrm{q}(x-\Phi, \Theta) \right\rrbracket.
\end{align}

\vspace{0.5em}
\end{widetext}

\section{CALCULATION OF THE TRAVELING VELOCITY}
\label{appendix:B}

\par 
In this appendix, we describe the method for calculating the nonzero traveling velocity $c$ of a traveling breather.
We assume that the system size is sufficiently large and that the period of $\Theta$ is shorter than that of $\Phi$, i.e., $2\pi/ \omega < 2L/ c$.
We consider the unperturbed solution $\boldsymbol{X}(x, t) = \boldsymbol{X}_0(x - \Phi(t), \Theta(t))$, which is obtained by solving Eq.~(\ref{eq:RD-systems}). 
For notational simplicity, we consider the case of two components ($N = 2$), as in the FHN and Gray--Scott models.
That is,
\begin{align}
\boldsymbol{X}_0(x - \Phi, \Theta) = \begin{pmatrix} u_0(x - \Phi, \Theta) \\ v_0(x - \Phi, \Theta) \end{pmatrix}.
\end{align}

\par 
Let us define the following periodic function:
\begin{align}
\label{eq:value_A}
& A\left( \Phi, \Theta \right) 
:= \int_0^{2L} \mathrm{d}x\, u_0(x - \Phi, \Theta) \exp\left[ i \pi \frac{x}{L} \right]
\in \mathbb{C}.
\end{align}
From the definition of $A(\Phi, \Theta)$, we obtain the following expression:
\begin{align}
\label{eq:argA}
\arg A(\Phi, \Theta)
= \frac{\pi}{L} \left( \Phi + B(\Theta) \right),
\end{align}
where $B(\Theta) \in \mathbb{R}$ is a $2\pi$-periodic function.  
Note that it is not necessary to compute the waveform of $B(\Theta)$ in the evaluation of the traveling velocity.
By substituting $t_{n-1} = (n - 1)T_{\Theta}$ and $t_n = n T_{\Theta}$ into Eq.~(\ref{eq:argA}) and taking the difference between the two resulting expressions, we obtain the following identity, which holds for any integer $n$:
\begin{align}
\label{eq:identity_for_c}
\frac{L}{\pi} \left[
\arg A(\Phi(t_n), \Theta(t_n)) - \arg A(\Phi(t_{n-1}), \Theta(t_{n-1}))
\right]
= c T_{\Theta},
\end{align}
where $\Phi(t_n) = ct$ and $\Theta(t_n) = 2\pi n$.
Therefore, the traveling velocity $c$ can be evaluated using the values of $A(\Phi(t), \Theta(t))$ at $t = t_n$ and $t_{n-1}$.  
These values are calculated using the time series $\boldsymbol{X}(x, t) = \boldsymbol{X}_0(x - \Phi(t), \Theta(t))$, which is obtained by solving Eq.~(\ref{eq:RD-systems}).
Although in this study, we use the variable $u_0(x-\Phi, \Theta)$ to compute $A(\Phi, \Theta)$ (see Eq.~(\ref{eq:value_A})), other variables can also be used.

\section{CALCULATION OF THE SPATIAL AND TEMPORAL PHASES FROM TIME SERIES OBTAINED FROM DIRECT NUMERICAL SIMULATIONS}
\label{appendix:C}

\par 
In this appendix, we describe the method for calculating the spatial and temporal phases using the time series $\boldsymbol{X}_\sigma(x, t)$, which is obtained from the direct numerical simulations performed in Secs.~\ref{subsec:coupled_FHN} and \ref{subsec:coupled_Gray--Scott_Model}.
For convenience, we omit subscript $\sigma$ and denote $\boldsymbol{X}_\sigma(x, t)$, $\Phi_\sigma(t)$, and $\Theta_\sigma(t)$ simply as $\boldsymbol{X}(x, t)$, $\Phi(t)$, and $\Theta(t)$, respectively.

\par 
Note that we distinguish between the perturbed solution of Eq.~(\ref{eq:coupled_RDsystems}), denoted by $\boldsymbol{X}(x, t)$, and the unperturbed solution of Eq.~(\ref{eq:RD-systems_with_gradient}), denoted by $\boldsymbol{X}_0(x - \Phi, \Theta)$.
We assume that the system size is sufficiently large compared to the size of the standing or traveling breather.
For notational simplicity, we consider two components ($N = 2$) similar to the case of the FHN and Gray--Scott models. 
That is,
\begin{align}    
\boldsymbol{X}(x, t) 
= \begin{pmatrix} 
    u(x, t) \\ v(x, t) 
\end{pmatrix},
\\ 
\boldsymbol{X}_0(x - \Phi, \Theta) 
= \begin{pmatrix} 
    u_0(x - \Phi, \Theta) \\ v_0(x - \Phi, \Theta) 
\end{pmatrix}.
\end{align}
This assumption is made only for notational simplicity, and the method is applicable to systems with any number of components (i.e., any $N$).
Furthermore, we assume that the coupling intensity is sufficiently weak ($\varepsilon \ll 1$) so that a solution $\boldsymbol{X}(x,t)$ remains in the vicinity of the orbit of the limit-torus solution $\boldsymbol{X}_0(x-\Phi, \Theta)$.

\par
Initially, we calculate the following time series:
\begin{align}
\label{eq:X_bar}
\overline{\boldsymbol{X}}(t)
= \int_0^{2L} \mathrm{d}x\, \boldsymbol{X}(x, t).
\end{align}
We define the $2\pi$-periodic function based on the unperturbed solution
\begin{align}
\label{eq:X0_bar}
\overline{\boldsymbol{X}_0}(\Theta)
= \int_0^{2L} \mathrm{d}x\, \boldsymbol{X}_0(x-\Phi, \Theta).
\end{align}
This function depends only on the temporal phase $\Theta(t)$ because the integration over $2L$ eliminates the dependence on $\Phi(t)$.
To determine the temporal phase $\Theta(t)$ corresponding to $\boldsymbol{X}(x, t)$, we identify the value of $\Theta$ that minimizes the distance between $\overline{\boldsymbol{X}_0}(\Theta)$ and $\overline{\boldsymbol{X}}(t)$.
Figures~\ref{fig:fig15}(a) and \ref{fig:fig15}(b) show the waveforms of $\overline{\boldsymbol{X}_0}(\Theta)$ of the FHN model and Gray--Scott model, respectively.

\par
In the following, we describe methods for calculating the spatial phase in both the $c = 0$ and $c \neq 0$ cases.
For $c = 0$, we consider the limit-torus solution $\boldsymbol{X}_0(x - \Phi, \Theta)$ with $\Phi = 0$, which satisfies
\begin{align}
\label{eq:value_H}
& H(\Theta)
:= \int_0^{2L} \mathrm{d}x\, 
\left. u_0(x-\Phi, \Theta) \exp\left[ i \pi \frac{x}{L} \right] \right|_{\Phi=0}
\in \mathbb{R}^+.  
\end{align}
Furthermore, we consider the function $A(\Phi, \Theta)$, which is defined in Eq.~(\ref{eq:value_A}).  
For convenience, we restate it below:
\begin{align*}
& A\left( \Phi, \Theta \right) 
:= \int_0^{2L} \mathrm{d}x\, u_0(x - \Phi, \Theta) \exp\left[ i \pi \frac{x}{L} \right]
\in \mathbb{C}.
\end{align*}
Note that the expression for $A(\Phi, \Theta)$ is identical to that of $H(\Theta)$, except that the spatial phase $\Phi$ is allowed to take arbitrary values.
From the definitions of $H(\Theta)$ and $A(\Phi, \Theta)$, we obtain the following relation:
\begin{align}
\label{eq:relation_H&A}
A(\Phi, \Theta) = H(\Theta) \exp\left( i \pi \frac{\Phi}{L} \right).
\end{align}
We approximate $A(\Phi, \Theta)$ as
\begin{align}
\label{eq:value_tildeA}
& \tilde{A}(t)
:= \int_0^{2L} \mathrm{d}x\, u(x, t) \exp\left[ i \pi \frac{x}{L} \right]
\in \mathbb{C}.
\end{align}
Thus, the spatial phase is determined as follows:
\begin{align}
\label{eq:Phi(c=0)}
\Phi(t) = \frac{L}{\pi} \arg \tilde{A}(t).
\end{align}

\par
\begin{figure*}[h]
    \begin{center}
        \includegraphics[scale=0.9]{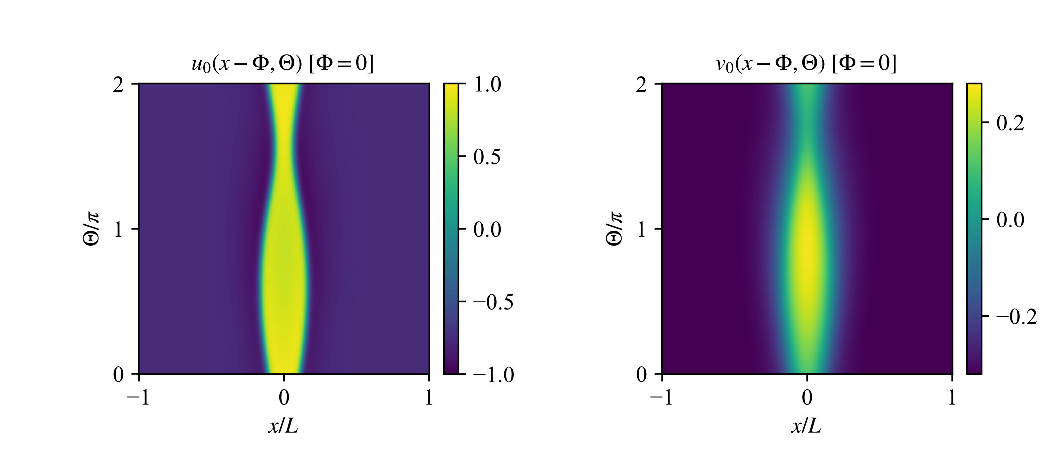}
        \caption{
        Limit-torus solution, 
        $\boldsymbol{X}_0(x-\Phi, \Theta) 
        = (u_0(x-\Phi, \Theta), v_0(x-\Phi, \Theta))$ 
        with $\Phi=0$, for the FHN model.
        }
        \label{fig:fig2}
    \end{center}
\end{figure*}
\begin{figure*}[h]
    \begin{center}
        \includegraphics[scale=0.9]{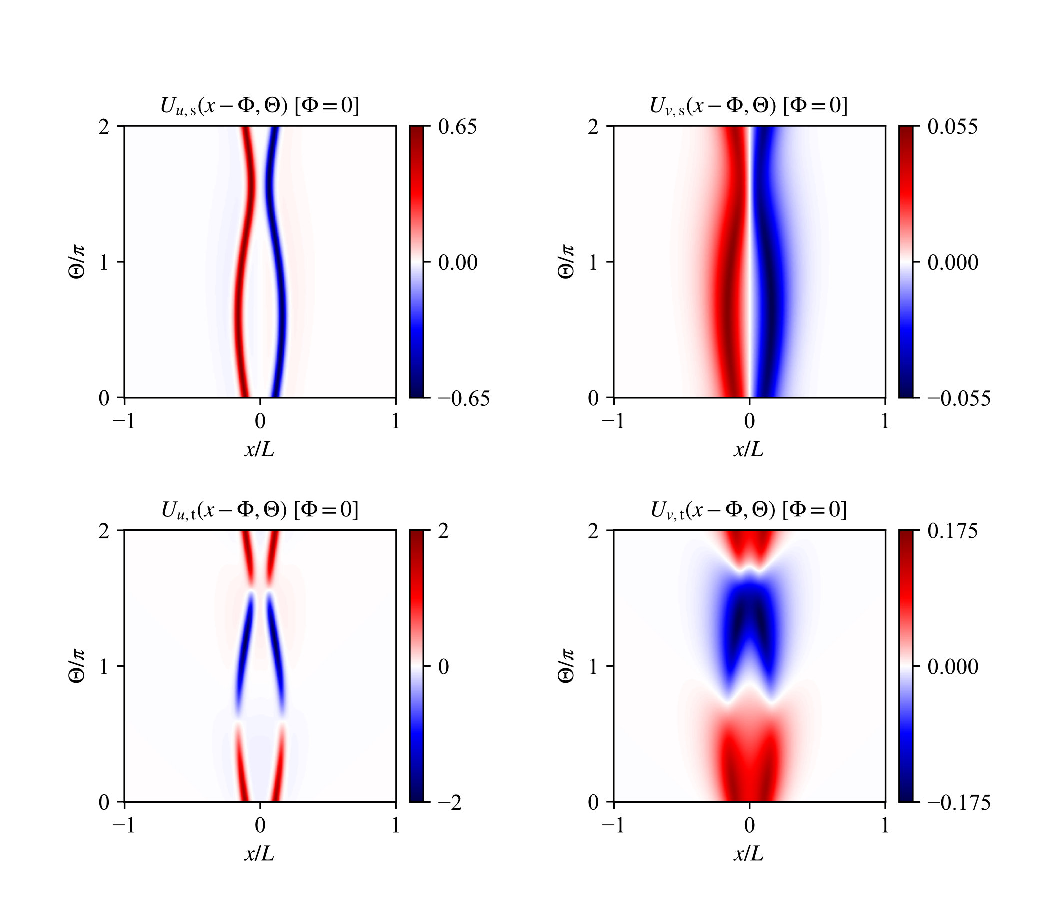}
        \caption{
        Floquet zero eigenfunctions, 
        $\boldsymbol{U}_\mathrm{s}(x - \Phi, \Theta) 
        = (U_{u,\mathrm{s}}(x - \Phi, \Theta), U_{v,\mathrm{s}}(x - \Phi, \Theta))$ 
        and 
        $\boldsymbol{U}_\mathrm{t}(x - \Phi, \Theta)
        = (U_{u,\mathrm{t}}(x - \Phi, \Theta), U_{v,\mathrm{t}}(x - \Phi, \Theta))$  with $\Phi=0$, for the FHN model.
        }
        \label{fig:fig3}
    \end{center}
\end{figure*}
\begin{figure*}[h]
    \begin{center}
        \includegraphics[scale=0.9]{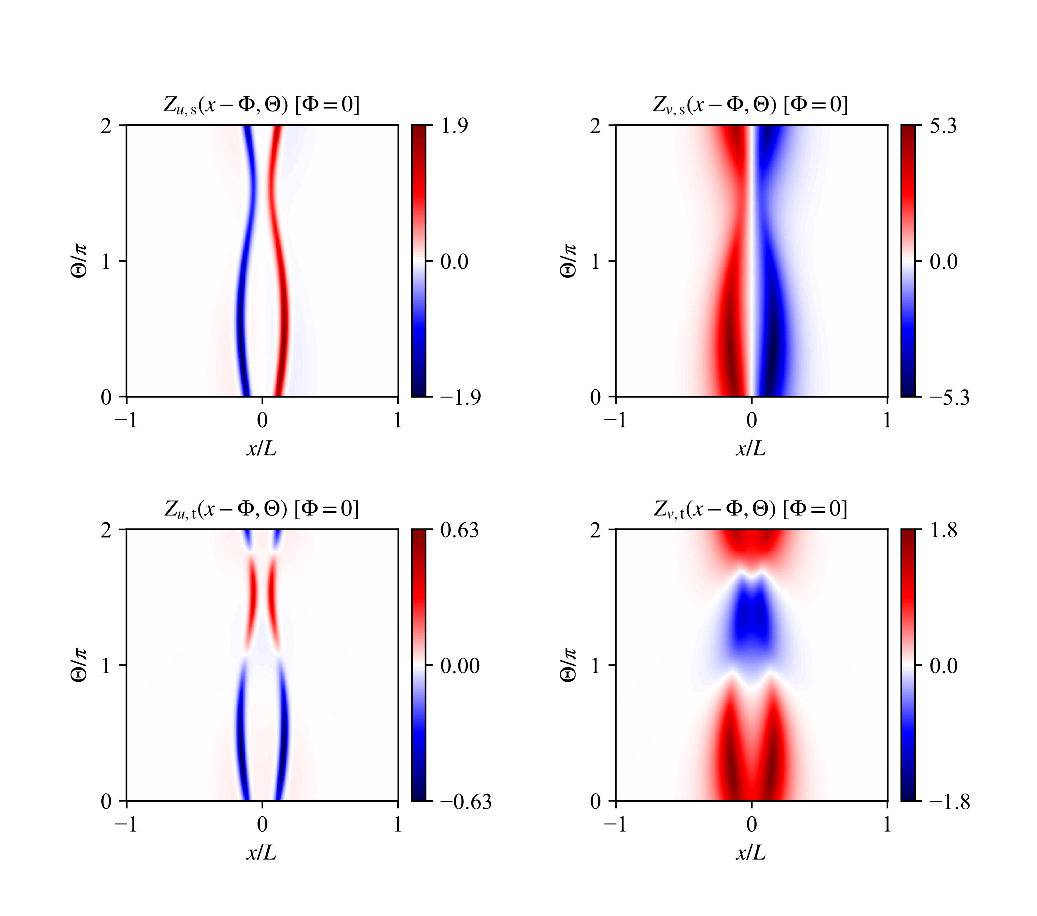}
        \caption{
        Phase sensitivity functions, 
        $\boldsymbol{Z}_\mathrm{s}(x - \Phi, \Theta) 
        = (Z_{u,\mathrm{s}}(x - \Phi, \Theta), Z_{v,\mathrm{s}}(x - \Phi, \Theta))$ 
        and 
        $\boldsymbol{Z}_\mathrm{t}(x - \Phi, \Theta)
        = (Z_{u,\mathrm{t}}(x - \Phi, \Theta), Z_{v,\mathrm{t}}(x - \Phi, \Theta))$ 
        with $\Phi=0$, for the FHN model.
        }
        \label{fig:fig4}
    \end{center}
\end{figure*}
\begin{figure*}[h]
    \begin{center}
        \includegraphics[scale=0.9]{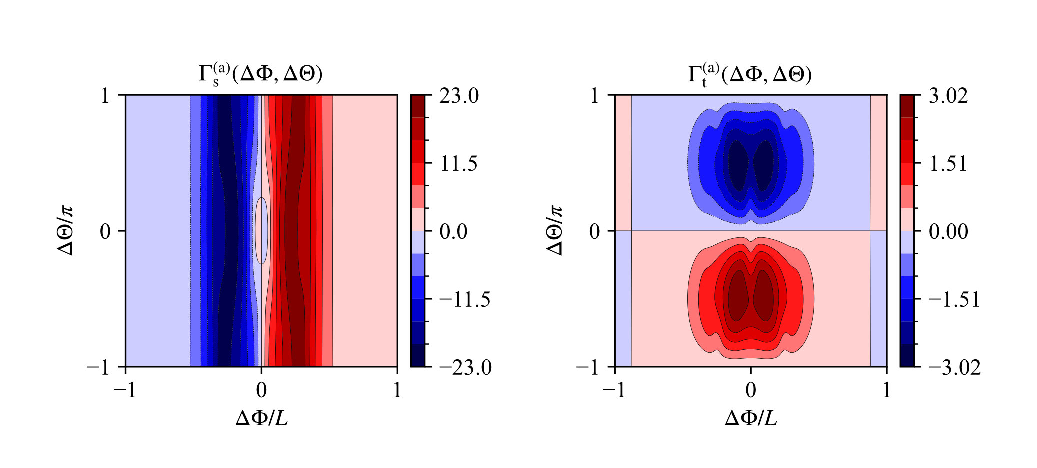}
        \caption{
        Antisymmetric components of the phase coupling functions, $\Gamma_\mathrm{s}^{(\mathrm{a})}(\Delta \Phi, \Delta \Theta)$ and $\Gamma_\mathrm{t}^{(\mathrm{a})}(\Delta \Phi, \Delta \Theta)$, for the coupled FHN models with K=$\mathrm{diag}(1,1)$.
        }
        \label{fig:fig5}
    \end{center}
\end{figure*}
\begin{figure*}[h]
    \begin{center}
        \includegraphics[scale=0.9]{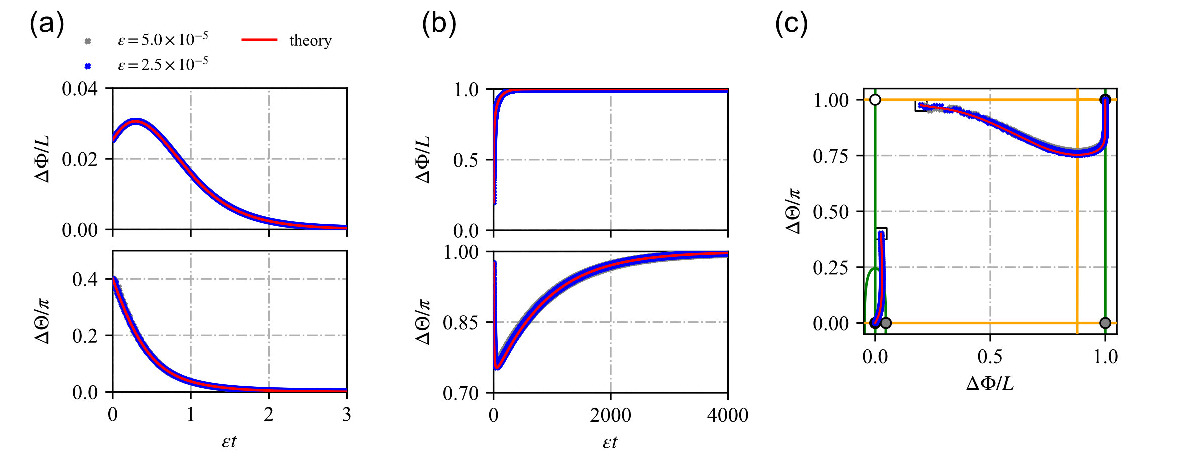}
        \caption{
         Comparison between direct numerical simulations [$\varepsilon = 5.0 \times 10^{-5}$ (gray) and $2.5 \times 10^{-5}$ (blue)] and theoretical values (red) for the coupled FHN models with ${\rm K} = \mathrm{diag}(1,1)$.
        (a)~
        Time evolution of the phase differences, $\Delta \Phi/L$ and $\Delta \Theta/\pi$, as functions of $\varepsilon t$, starting from $(\Delta \Phi/L, \Delta \Theta/\pi) \simeq (0.025, 0.4)$.  
        (b)~
        Same as panel (a) but starting from $(\Delta \Phi/L, \Delta \Theta/\pi) \simeq (0.2, 0.975)$.  
        (c)~
        Trajectories in the $(\Delta \Phi/L, \Delta \Theta/\pi)$ space corresponding to the time evolutions shown in panels (a) and (b).
        The green and orange lines indicate the nullclines of $\Gamma_\mathrm{s}^{(\mathrm{a})}(\Delta \Phi, \Delta \Theta)$ and $\Gamma_\mathrm{t}^{(\mathrm{a})}(\Delta \Phi, \Delta \Theta)$, respectively.
        Stable and unstable fixed points are indicated by closed and open circles, respectively; saddle points are indicated by gray circles. 
        Considering Eqs.~(\ref{eq:antisymmertry(point)_Gamma_s^a}), (\ref{eq:antisymmertry(point)_Gamma_t^a}), and (\ref{eq:Gamma_s^a_symmetry})--(\ref{eq:Gamma_t^a_symmetry2}), the numbers of stable fixed points, unstable fixed points, and saddle points within the rectangular region defined by $-1 < \Delta\Phi/L \leq 1$ and $-1 < \Delta\Theta/\pi \leq 1$ are 2, 1, and 3, respectively.
        The initial conditions are indicated by open squares.
        }
        \label{fig:fig6}
    \end{center}
\end{figure*}
\begin{figure*}[h]
    \begin{center}
        \includegraphics[scale=0.9]{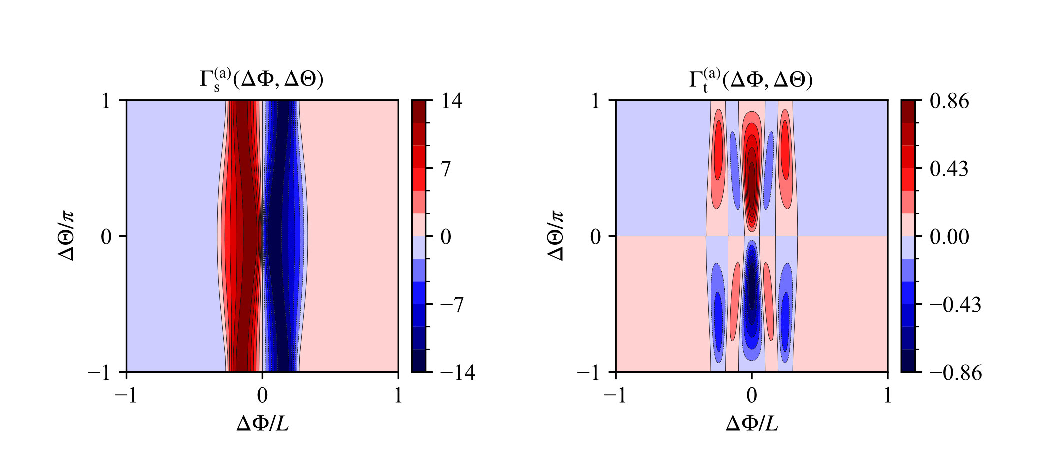}
        \caption{
        Antisymmetric components of the phase coupling functions, $\Gamma_\mathrm{s}^{(\mathrm{a})}(\Delta \Phi, \Delta \Theta)$ and $\Gamma_\mathrm{t}^{(\mathrm{a})}(\Delta \Phi, \Delta \Theta)$, for the coupled FHN models with K=$\mathrm{diag}(1,0)$.
        }
        \label{fig:fig7}
    \end{center}
\end{figure*}
\begin{figure*}[h]
    \begin{center}
        \includegraphics[scale=0.9]{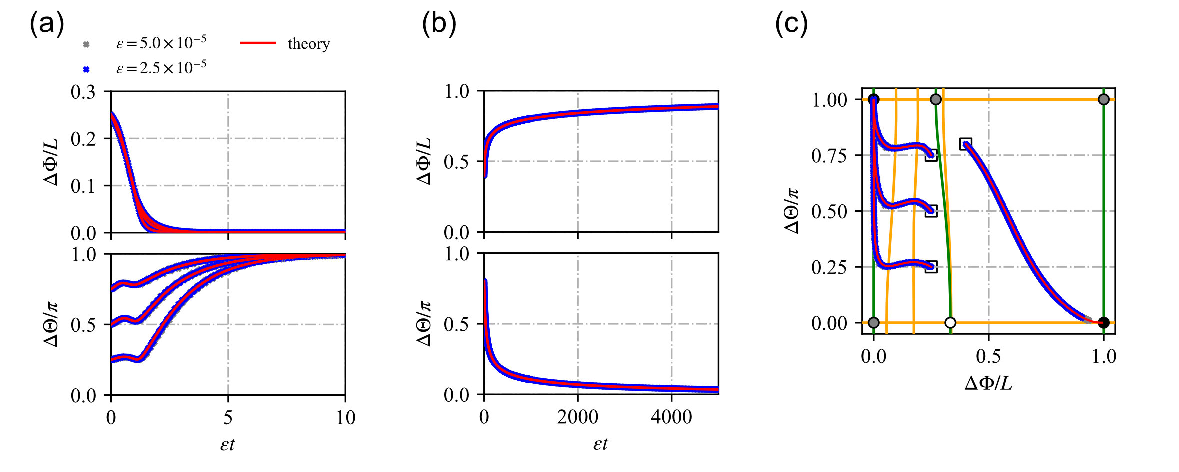}
        \caption{
         Comparison between direct numerical simulations [$\varepsilon = 5.0 \times 10^{-5}$ (gray) and $2.5 \times 10^{-5}$ (blue)] and theoretical values (red) for the coupled FHN models with ${\rm K} = \mathrm{diag}(1,0)$.
        (a)~
        Time evolution of the phase differences, $\Delta \Phi/L$ and $\Delta \Theta/\pi$, as functions of $\varepsilon t$, starting from $(\Delta \Phi/L, \Delta \Theta/\pi) \simeq (0.25, 0.25)$, $(0.25, 0.50)$, and $(0.25, 0.75)$.  
        (b)~
        Same as panel (a) but starting from $(\Delta \Phi/L, \Delta \Theta/\pi) \simeq (0.4, 0.8)$.  
        (c)~
        Trajectories in the $(\Delta \Phi/L, \Delta \Theta/\pi)$ space corresponding to the time evolutions shown in panels (a) and (b).
        The green and orange lines indicate the nullclines of $\Gamma_\mathrm{s}^{(\mathrm{a})}(\Delta \Phi, \Delta \Theta)$ and $\Gamma_\mathrm{t}^{(\mathrm{a})}(\Delta \Phi, \Delta \Theta)$, respectively.
        Stable and unstable fixed points are indicated by closed and open circles, respectively; saddle points are indicated by gray circles. 
        Considering Eqs.~(\ref{eq:antisymmertry(point)_Gamma_s^a}), (\ref{eq:antisymmertry(point)_Gamma_t^a}), and (\ref{eq:Gamma_s^a_symmetry})--(\ref{eq:Gamma_t^a_symmetry2}), the numbers of stable fixed points, unstable fixed points, and saddle points within the rectangular region defined by $-1 < \Delta\Phi/L \leq 1$ and $-1 < \Delta\Theta/\pi \leq 1$ are 2, 2, and 4, respectively.
        The initial conditions are indicated by open squares.
        One of the trajectories shown in panel~(c), corresponding to the trajectory in panel~(b), was terminated before reaching the fixed point because of the limitations in computational time. 
        The theoretical value (red) for this trajectory was computed up to $\varepsilon t = 9 \times 10^4$.
        The range of $\varepsilon t$ shown in panel~(b) is limited to $5000$.
        }
        \label{fig:fig8}
    \end{center}
\end{figure*}
\begin{figure}[h]
    \begin{center}
        \includegraphics[scale=0.9]{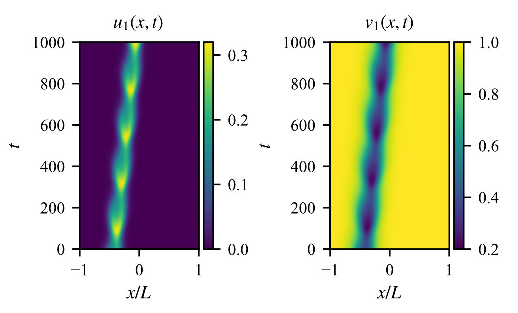}
        \caption{
        Example of time series 
        $\boldsymbol{X}_1(x,t) = (u_1(x,t), v_1(x,t))$ 
        of the Gray--Scott model in the absence of coupling ($\varepsilon = 0$).
        }
        \label{fig:fig9}
    \end{center}
\end{figure}
\begin{figure*}[h]
    \begin{center}
        \includegraphics[scale=0.9]{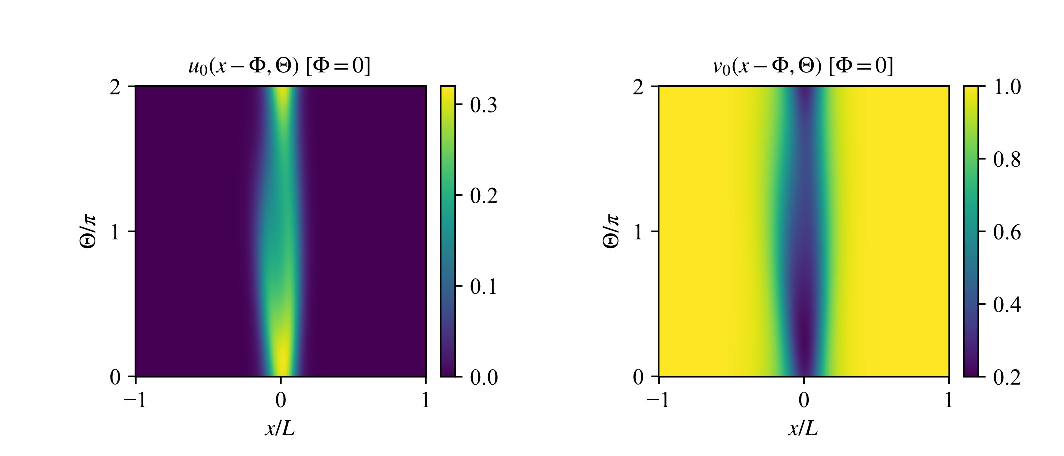}
        \caption{
        Limit-torus solution, 
        $\boldsymbol{X}_0(x-\Phi, \Theta) 
        = (u_0(x-\Phi, \Theta), v_0(x-\Phi, \Theta))$ 
        with $\Phi=0$, for the Gray--Scott model.
        }
        \label{fig:fig10}
    \end{center}
\end{figure*}
\begin{figure*}[h]
    \begin{center}
        \includegraphics[scale=0.9]{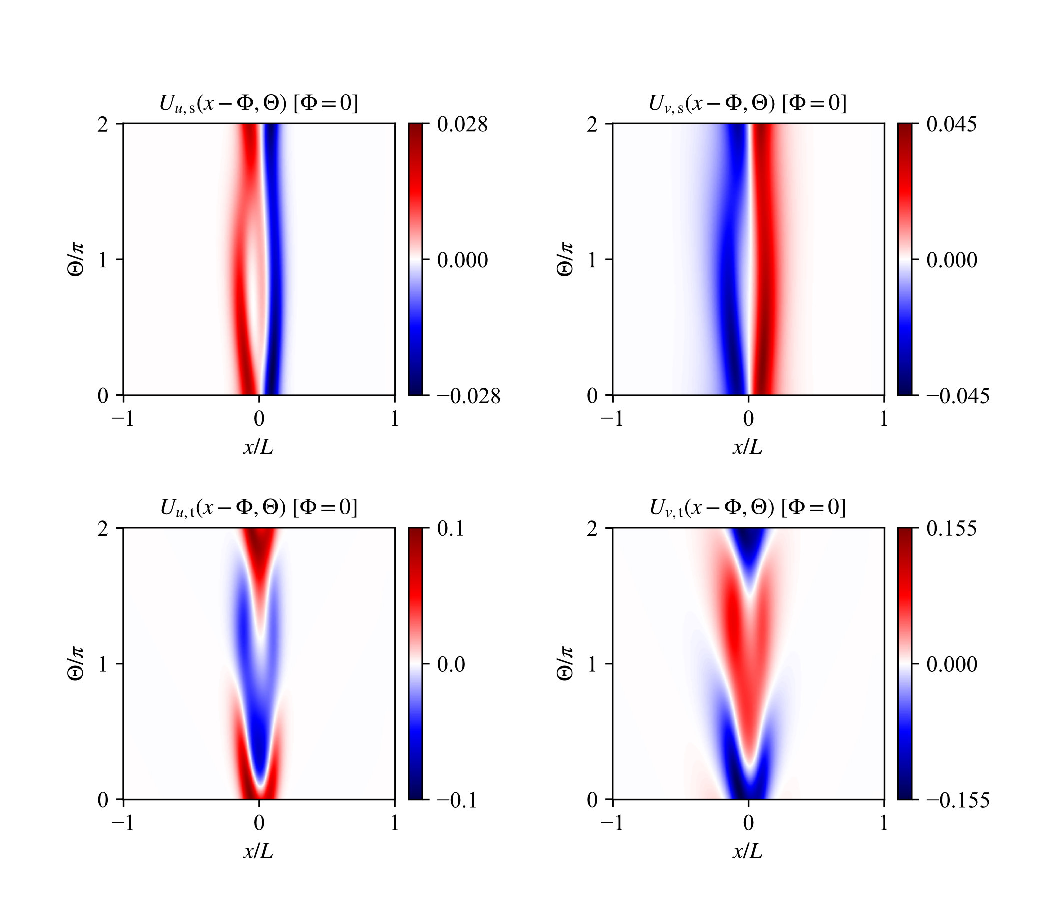}
        \caption{
        Floquet zero eigenfunctions, 
        $\boldsymbol{U}_\mathrm{s}(x - \Phi, \Theta) 
        = (U_{u,\mathrm{s}}(x - \Phi, \Theta), U_{v,\mathrm{s}}(x - \Phi, \Theta))$ 
        and 
        $\boldsymbol{U}_\mathrm{t}(x - \Phi, \Theta)
        = (U_{u,\mathrm{t}}(x - \Phi, \Theta), U_{v,\mathrm{t}}(x - \Phi, \Theta))$ 
        with $\Phi=0$, for the Gray--Scott model.
        }
        \label{fig:fig11}
    \end{center}
\end{figure*}
\begin{figure*}[h]
    \begin{center}
        \includegraphics[scale=0.9]{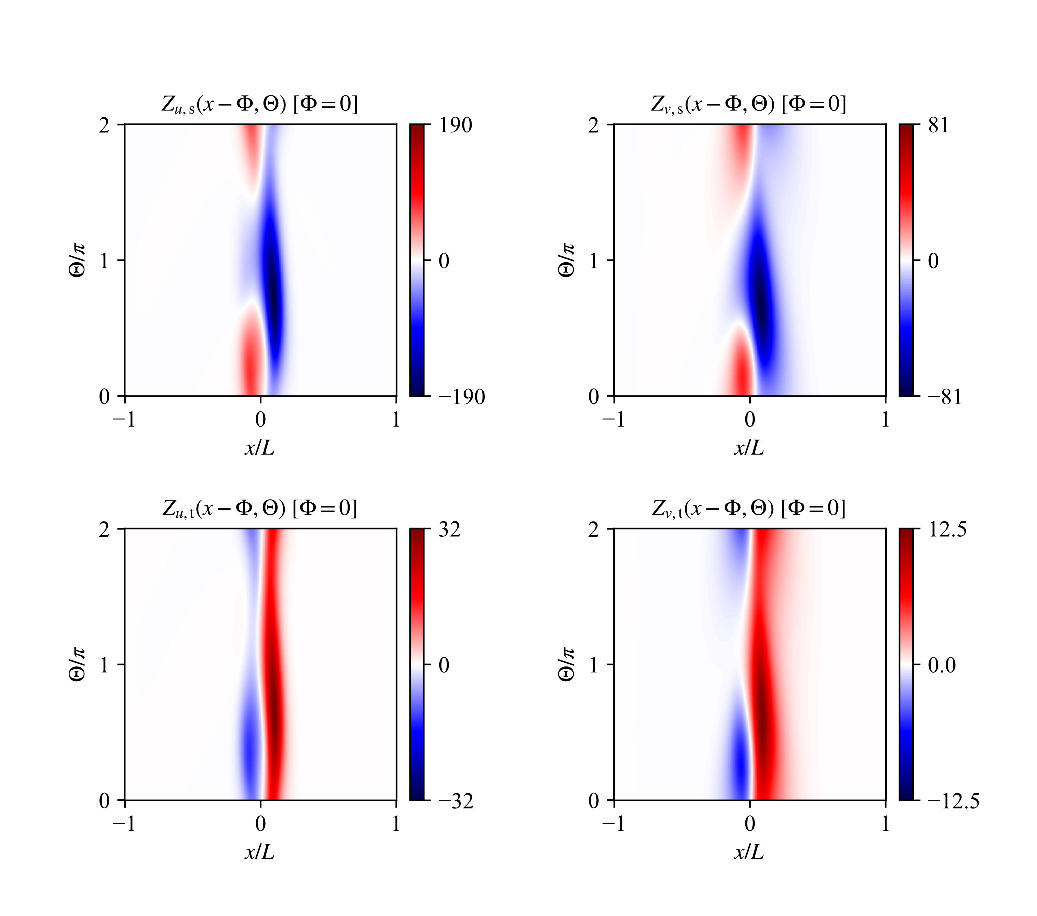}
        \caption{
        Phase sensitivity functions, 
        $\boldsymbol{Z}_\mathrm{s}(x - \Phi, \Theta) 
        = (Z_{u,\mathrm{s}}(x - \Phi, \Theta), Z_{v,\mathrm{s}}(x - \Phi, \Theta))$ 
        and 
        $\boldsymbol{Z}_\mathrm{t}(x - \Phi, \Theta)
        = (Z_{u,\mathrm{t}}(x - \Phi, \Theta), Z_{v,\mathrm{t}}(x - \Phi, \Theta))$ 
        with $\Phi=0$, for the Gray--Scott model.
        }
        \label{fig:fig12}
    \end{center}
\end{figure*}
\begin{figure*}[h]
    \begin{center}
        \includegraphics[scale=0.9]{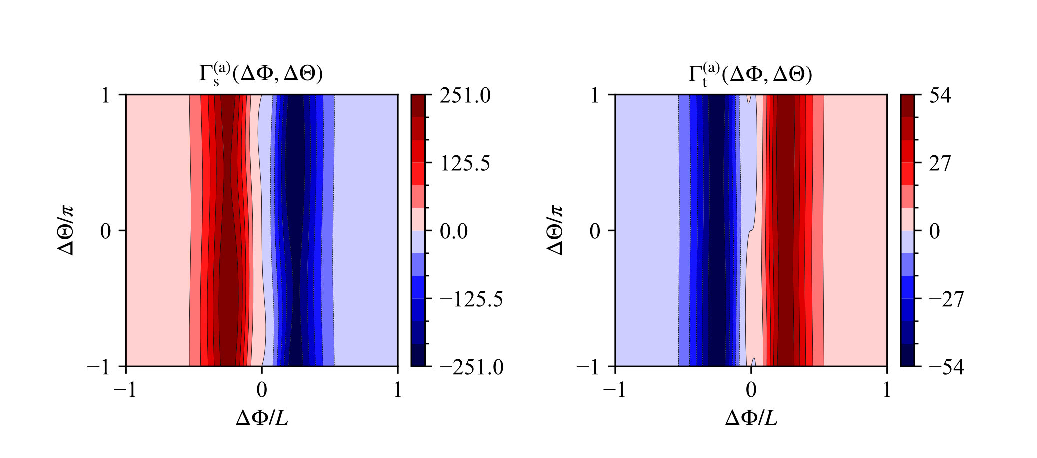}
        \caption{
        Antisymmetric components of the phase coupling functions, $\Gamma_\mathrm{s}^{(\mathrm{a})}(\Delta \Phi, \Delta \Theta)$ and $\Gamma_\mathrm{t}^{(\mathrm{a})}(\Delta \Phi, \Delta \Theta)$, for the coupled Gray--Scott models.
        }
        \label{fig:fig13}
    \end{center}
\end{figure*}
\begin{figure*}[h]
    \begin{center}
        \includegraphics[scale=0.85]{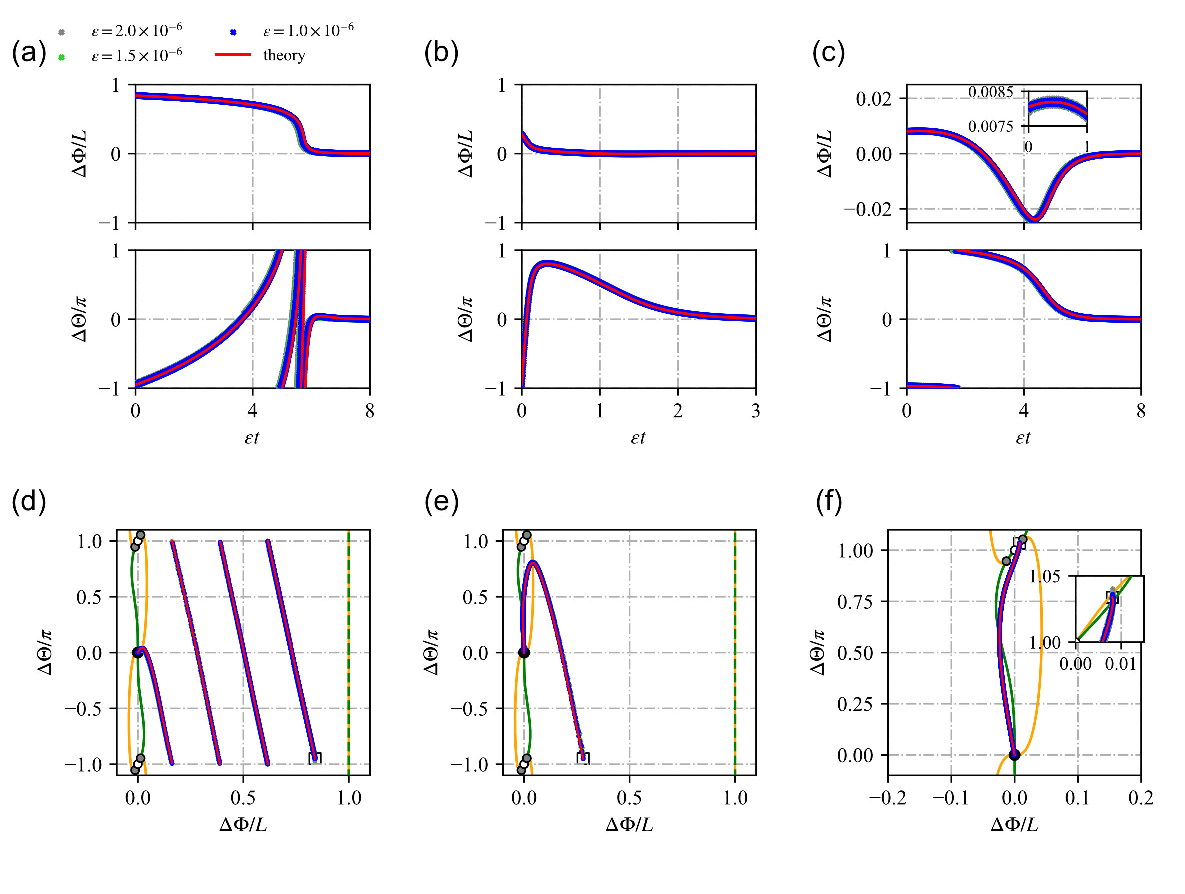}
        \caption{
        Comparison between direct numerical simulations [$\varepsilon = 2.0 \times 10^{-6}$ (gray), $1.5 \times 10^{-6}$ (green), and $1.0 \times 10^{-6}$ (blue)] and theoretical values (red) for the coupled Gray--Scott models.
        (a)--(c)~
        Time evolution of the phase differences, $\Delta \Phi/L$ and $\Delta \Theta/\pi$, as functions of $\varepsilon t$, starting from $(\Delta \Phi/L, \Delta \Theta/\pi) \simeq (0.84, -0.95)$, $(0.28, -0.95)$, and $(0.008, -0.96629)$. 
        (d)--(f)~
        Trajectories in the $(\Delta \Phi/L, \Delta \Theta/\pi)$ space corresponding to panels (a)--(c). 
        The green and orange lines indicate the nullclines of $\Gamma_\mathrm{s}^{(\mathrm{a})}(\Delta \Phi, \Delta \Theta)$ and $\Gamma_\mathrm{t}^{(\mathrm{a})}(\Delta \Phi, \Delta \Theta)$, respectively.
        Stable and unstable fixed points are indicated by closed and open circles, respectively; saddle points are indicated by gray circles. 
        The numbers of stable fixed points, unstable fixed points, and saddle points within the rectangular region defined by $-1 < \Delta\Phi/L \leq 1$ and $-1 < \Delta\Theta/\pi \leq 1$ are 1, 1, and 2, respectively.
        Furthermore, the nullclines of $\Gamma_\mathrm{s}^{(\mathrm{a})}(\Delta \Phi, \Delta \Theta)$ and $\Gamma_\mathrm{t}^{(\mathrm{a})}(\Delta \Phi, \Delta \Theta)$ overlap at $\Delta \Phi / L = \pm 1$, where the green line is depicted as a broken line.
        The initial conditions are indicated by open squares.
        The insets in panels (c) and (f) show the early-time behavior, where the spatial phase difference initially increases and then decreases.
        }
        \label{fig:fig14}
    \end{center}
\end{figure*}
\begin{figure}[h]
    \begin{center}
        \includegraphics[scale=0.9]{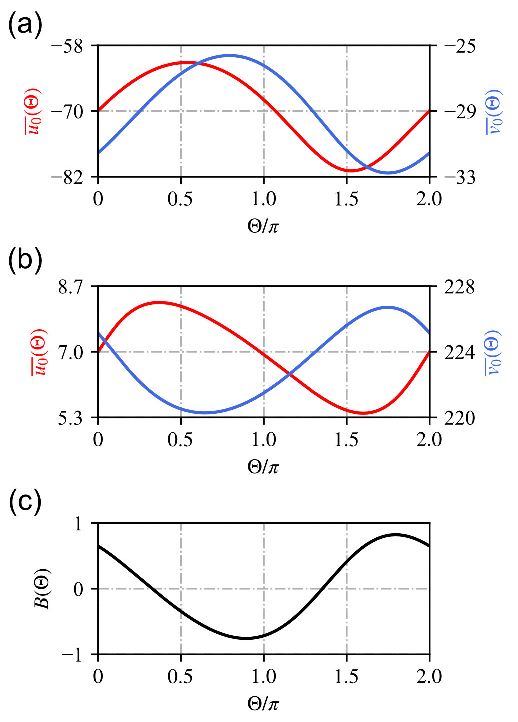}
        \caption{
        Waveforms of the periodic functions $\overline{\boldsymbol{X}_0}(\Theta)$ and $B(\Theta)$.  
        (a)
        $\overline{\boldsymbol{X}_0}(\Theta)$ for the FHN model.
        (b)
        $\overline{\boldsymbol{X}_0}(\Theta)$ for the Gray--Scott model.
        (c)~
        $B(\Theta)$ for the Gray--Scott model.
        }
        \label{fig:fig15}
    \end{center}
\end{figure}

Next, we describe the method for calculating the spatial phase for $ c \neq 0 $. 
We consider function $B(\Theta)$ in Eq.~(\ref{eq:argA}), which satisfies the following condition:
\begin{align}
\label{eq:condition_funcB}
\frac{1}{2\pi} \int_0^{2\pi} \mathrm{d}\Theta \, B(\Theta) = 0.
\end{align}
The above equation means that the constant term in $B(\Theta)$ is zero.
By setting $\Phi = 0$ in Eq.~(\ref{eq:argA}), the periodic function \(B(\Theta)\) is determined as follows:
\begin{align}
\label{eq:function_B}
& B(\Theta) = \frac{L}{\pi} \arg A(0, \Theta).
\end{align}
The waveform of $B(\Theta)$, computed using the limit-torus solution of the Gray--Scott model, is shown in Fig.~\ref{fig:fig15}(c).
Given Eq.~(\ref{eq:function_B}), the condition described in Eq.~(\ref{eq:condition_funcB}) ensures that the state corresponding to $\Phi = 0$ in the limit-torus solution $\bm{X}_0(x - \Phi, \Theta)$ is uniquely determined.
To obtain the time series of the spatial phase, we approximate $A(\Phi, \Theta)$ in Eq.~(\ref{eq:argA}) by $\tilde{A}(t)$.
Thus, the spatial phase $\Phi(t)$ corresponding to $\boldsymbol{X}(x,t)$ is determined by
\begin{align}
\label{eq:Phi(cneq0)}
\Phi(t) = \frac{L}{\pi} \arg \tilde{A}(t) - B(\Theta(t)).
\end{align}
It can be readily seen that Eq.~(\ref{eq:Phi(c=0)}) is a special case of Eq.~(\ref{eq:Phi(cneq0)}); when $c = 0$, $B(\Theta)$ is zero because of the spatial reflection symmetry (see Eq.~(\ref{eq:reflection_symmetry_X0_FHN}) and Fig.~\ref{fig:fig2}).
Although we use the variables $u(x, t)$ and $u_0(x - \Phi, \Theta)$ to compute $\tilde{A}(t)$ and $A(\Phi, \Theta)$, respectively, other variables can also be used.


%

\end{document}